\documentclass[11pt, a4paper]{article}

\makeatletter%

\usepackage{ifthen}



\makeatletter%

\makeatletter
\newcommand{\ifndef}[2]{\@ifundefined{#1}{#2}{}}
\makeatother

\newcommand{\mydef}[2]{\def#1{#2}}

\newcommand{\nospell}[1]{#1}  %

\makeatletter
\newcommand{\myusepackage}[2][]{\@ifpackageloaded{#2}{} %
{\ifthenelse{\equal{}{#1}} {\usepackage{#2}} {\usepackage[#1]{#2}} }}
\makeatother

\myusepackage{amssymb}
\myusepackage{amsmath}  %

\myusepackage{amsthm}  %

{}
\myusepackage[T1,OT2,T2A]{fontenc}
\DeclareTextSymbolDefault{\CYRYAT}{OT2}
\DeclareTextSymbolDefault{\cyryat}{OT2}
\DeclareTextSymbolDefault{\CYRFITA}{OT2}
\DeclareTextSymbolDefault{\cyrfita}{OT2}
\DeclareTextSymbolDefault{\CYRIZH}{OT2}
\DeclareTextSymbolDefault{\cyrizh}{OT2}
\myusepackage[utf8]{inputenc}
\let\f\relax

{}
\usepackage[greek,ukrainian,russian,english]{babel}
{}

\myusepackage{latexsym}
\myusepackage{amsfonts}
\myusepackage{units}    %
\myusepackage{psfrag}   %
\myusepackage{url}   %
\myusepackage{hyperref}  %
\myusepackage{hyphenat}  %
\myusepackage{caption}   %
\myusepackage{microtype}   %
\myusepackage{marginnote}  %
\myusepackage{calc}  %
{}  %
\myusepackage{enumitem}  %
{}
\myusepackage[dvips]{graphicx}
\myusepackage[usenames,dvipsnames]{color}
{}
{}  %
{} %

{}     %
\newcommand{\dgCapDefinition}{Definition}
\newcommand{\dgCapDefinitions}{Definitions}
\newcommand{\dgCapPostulate}{Postulate}
\newcommand{\dgCapPostulates}{Postulates}
\newcommand{\dgCapExample}{Example}

\newcommand{\dgCapFact}{Fact}
\newcommand{\dgCapFacts}{Facts}
\newcommand{\dgCapQuestion}{Question}
\newcommand{\dgCapQuestions}{Questions}
\newcommand{\dgCapLemma}{Lemma}
\newcommand{\dgCapLemmas}{Lemmas}

\newcommand{\dgCapCorollary}{Corollary}
\newcommand{\dgCapCorollaries}{Corollaries}
\newcommand{\dgCapProposition}{Proposition}
\newcommand{\dgCapPropositions}{Propositions}
\newcommand{\dgCapClaim}{Claim}
\newcommand{\dgCapClaims}{Claims}
\newcommand{\dgCapTheorem}{Theorem}
\newcommand{\dgCapTheorems}{Theorems}
\newcommand{\dgCapProblem}{Problem}
\newcommand{\dgCapProblems}{Problems}
\newcommand{\dgCapRemark}{Remark}
\newcommand{\dgCapRemarks}{Remarks}
\newcommand{\dgCapConjecture}{Conjecture}
\newcommand{\dgCapConjectures}{Conjectures}
\newcommand{\dgCapResult}{Result}

\newcommand{\dgCapChapter}{Chapter}
\newcommand{\dgCapChapters}{Chapters}
\newcommand{\dgCapSection}{Section}
\newcommand{\dgCapSections}{Sections}
\newcommand{\dgCapSubsection}{Subsection}
\newcommand{\dgCapSubsections}{Subsections}
\newcommand{\dgCapFigure}{Figure}
\newcommand{\dgCapFigures}{Figures}
\newcommand{\dgCapEquation}{Equation}
\newcommand{\dgCapEquations}{Equations}
\newcommand{\dgCapExpression}{Expression}
\newcommand{\dgCapExpressions}{Expressions}
\newcommand{\dgCapInequality}{Inequality}
\newcommand{\dgCapInequalities}{Inequalities}
\newcommand{\dgProofOf}{\proofname\ of}
{}
\newcommand{\dgDefinition}{Definition}
\newcommand{\dgDefinitions}{Definitions}
\newcommand{\dgPostulate}{Postulate}
\newcommand{\dgPostulates}{Postulates}

\newcommand{\dgFact}{Fact}
\newcommand{\dgFacts}{Facts}
\newcommand{\dgQuestion}{Question}
\newcommand{\dgQuestions}{Questions}
\newcommand{\dgLemma}{Lemma}
\newcommand{\dgLemmas}{Lemmas}
\newcommand{\dgCorollary}{Corollary}
\newcommand{\dgCorollaries}{Corollaries}
\newcommand{\dgProposition}{Proposition}
\newcommand{\dgPropositions}{Propositions}
\newcommand{\dgClaim}{Claim}
\newcommand{\dgClaims}{Claims}
\newcommand{\dgTheorem}{Theorem}
\newcommand{\dgTheorems}{Theorems}
\newcommand{\dgProblem}{Problem}
\newcommand{\dgProblems}{Problems}
\newcommand{\dgRemark}{Remark}
\newcommand{\dgRemarks}{Remarks}
\newcommand{\dgConjecture}{Conjecture}
\newcommand{\dgConjectures}{Conjectures}

\newcommand{\dgChapter}{Chapter}
\newcommand{\dgChapters}{Chapters}
\newcommand{\dgSection}{Section}
\newcommand{\dgSections}{Sections}
\newcommand{\dgSubsection}{Subsection}
\newcommand{\dgSubsections}{Subsections}
\newcommand{\dgFigure}{Figure}
\newcommand{\dgFigures}{Figures}
\newcommand{\dgEquation}{Equation}
\newcommand{\dgEquations}{Equations}
\newcommand{\dgExpression}{Expression}
\newcommand{\dgExpressions}{Expressions}
\newcommand{\dgInequality}{Inequality}
\newcommand{\dgInequalities}{Inequalities}
{}
{}

{}
\ifndef{theorem}{}
\ifndef{theorem*}{\newtheorem*{theorem*}{\dgCapTheorem}}
{}

{}  %
{}  %
\ifndef{lemma}{}
\ifndef{lemma*}{\newtheorem*{lemma*}{\dgCapLemma}}
\ifndef{corollary}{}
\ifndef{corollary*}{\newtheorem*{corollary*}{\dgCapCorollary}}
\ifndef{conjecture}{}
\ifndef{conjecture*}{\newtheorem*{conjecture*}{\dgCapConjecture}}
\ifndef{proposition}{}
\ifndef{proposition*}{\newtheorem*{proposition*}{\dgCapProposition}}
\ifndef{claim}{}
\ifndef{claim*}{\newtheorem*{claim*}{\dgCapClaim}}
\ifndef{result}{}
\ifndef{result*}{\newtheorem*{result*}{\dgCapResult}}
\ifndef{problem}{}
\ifndef{problem*}{\newtheorem*{problem*}{\dgCapProblem}}
{}  %
{}  %

\newtheoremstyle{mydefinition}  %
{\topsep}{\topsep}  %
{\slshape}  %
{}  %
{\bfseries}  %
{.}  %
{ }  %
{}  %

\newtheoremstyle{myremark}  %
{\topsep}{\topsep}  %
{\slshape}  %
{}  %
{\bfseries\itshape}  %
{.}  %
{ }  %
{\thmname{#1}\thmnumber{ \!#2}}  %

\newtheoremstyle{myexample}  %
{\topsep}{\topsep}  %
{\itshape}  %
{}  %
{\slshape}  %
{.}  %
{ }  %
{\ul{\thmname{#1}}}  %

\newtheoremstyle{myclaims}  %
{\topsep}{\topsep}  %
{\slshape}  %
{}  %
{\bfseries\slshape}  %
{.}  %
{ }  %
{\thmname{#1}\thmnumber{ \!#2}\ifthenelse{\equal{}{#3}}%
{}{\textnormal{ \!(#3)}}}  %

{} %
{} %
{\theoremstyle{myremark}

\newtheorem*{myremark*}{\dgCapRemark}
}
{} %
{\theoremstyle{mydefinition}
\ifndef{postulate}{}
\ifndef{postulate*}{\newtheorem*{postulate*}{\dgCapPostulate}}
\ifndef{definition}{}
\ifndef{definition*}{\newtheorem*{definition*}{\dgCapDefinition}}
}
{\theoremstyle{myexample}
\ifndef{example}{}
\ifndef{example*}{\newtheorem*{example*}{\dgCapExample}}
}
{\theoremstyle{myclaims}

\newtheorem*{my_claim*}{\dgCapClaim}
\ifndef{fact}{}
\ifndef{fact*}{\newtheorem*{fact*}{\dgCapFact}}
\ifndef{question}{}
\ifndef{question*}{\newtheorem*{question*}{\dgCapQuestion}}
}
{} %
{} %
{} %

{}
\newtheoremstyle{anystatementst}  %
{\topsep}{\topsep}  %
{\itshape}  %
{}  %
{\bfseries}  %
{.}  %
{ }  %
{#3}  %

{\theoremstyle{anystatementst} }

{}

\newcommand{\MyUniPat}{lsdfgkhjvrkjlhmisdlcjn}

\newcommand{\newident}[3][\MyUniPat]{\ifthenelse{\equal{\MyUniPat}{#1}}
{
\newcommand{#2}[1][]{\Ensuremath{\mathit{#3##1}}}
}
{\ifthenelse{\equal{}{#1}}
{
\newcommand{#2}[1][]{\Ensuremath{\mathit{#3}}}
}
{
\newcommand{#2}[1][\MyUniPat]{\ifthenelse{\equal{\MyUniPat}{##1}}%
{\Ensuremath{\mathit{#1}}}%
{\Ensuremath{\mathit{#3}}}}
}
}
}

\newcommand{\newidenT}[3][\MyUniPat]{\ifthenelse{\equal{\MyUniPat}{#1}}
{
\newcommand{#2}[1][\MyUniPat]{\ifthenelse{\equal{\MyUniPat}{##1}}%
{\il{#3}}%
{\Ensuremath{\mathit{#3##1}}}}
}
{
\newcommand{#2}[1][\MyUniPat]{\ifthenelse{\equal{\MyUniPat}{##1}}%
{\il{#1}}%
{\Ensuremath{\mathit{#3}}}}
}
}

\newcommand{\newmat}[3][\MyUniPat]{\ifthenelse{\equal{\MyUniPat}{#1}}%
{\newcommand{#2}[1][]{\Ensuremath{#3##1}}}%
{\newcommand{#2}[1][]{\Ensuremath{#3}}}%
}

\newcommand{\providemat}[3][\MyUniPat]{\ifthenelse{\equal{\MyUniPat}{#1}}
{\providecommand{#2}[1][]{\Ensuremath{#3##1}}}
{\providecommand{#2}[1][]{\Ensuremath{#3}}}  %
}

\newcommand{\newmatop}[3][\MyUniPat]{\ifthenelse{\equal{\MyUniPat}{#1}}
{
\mydef{#2}{\operatorname{#3}}
}
{
\newcommand{#2}[1][\MyUniPat]{\ifthenelse{\equal{\MyUniPat}{##1}}%
{\operatorname{#1}}%
{\operatorname{#3}}}
}
}

\newcommand{\newfunction}[2]{%
\newcommand{#1}[2][\MyUniPat]{\ifthenelse{\equal{\MyUniPat}{##1}}%
{\Ensuremath{#2\lf(##2\rt)}}%
{#2(##2)}}%
}

\makeatletter
\newcommand{\MyMakeTheoMacros}[3]{
\expandafter\newcommand\csname\expandafter\@gobble\string#2NostarNoname@DGaux\endcsname[2][]
{\ifthenelse{\equal{}{##1}}%
{\begin{#1}~##2 \end{#1}}%
{\begin{#1}\label{##1}~##2\end{#1}}%
}
\expandafter\newcommand\csname\expandafter\@gobble\string#2StarNoname@DGaux\endcsname[1]
{\begin{#1*}~##1 \end{#1*}}
\def#2{\expandafter\@ifstar%
\expandafter{\csname\expandafter\@gobble\string#2StarNoname@DGaux\endcsname}%
{\csname\expandafter\@gobble\string#2NostarNoname@DGaux\endcsname}%
}

\expandafter\newcommand\csname\expandafter\@gobble\string#2NostarName@DGaux\endcsname[3][]
{\ifthenelse{\equal{}{##1}}%
{\begin{#1}[\e{##2}]~##3 \end{#1}}%
{\begin{#1}[\e{##2}]\label{##1}~##3\end{#1}}%
}
\expandafter\newcommand\csname\expandafter\@gobble\string#2StarName@DGaux\endcsname[2]
{\begin{#1*}[\e{##1}]~##2 \end{#1*}}
\def#3{\expandafter\@ifstar%
\expandafter{\csname\expandafter\@gobble\string#2StarName@DGaux\endcsname}
{\csname\expandafter\@gobble\string#2NostarName@DGaux\endcsname}%
}
}
\makeatother

\makeatletter
\newtheorem*{rep@theorem}{\rep@title}
\newcommand{\newreptheorem}[2]{%
\newenvironment{rep#1}[1]{%
\def\rep@title{#2 \ref{##1}}%
\begin{rep@theorem}}%
{\end{rep@theorem}}}
\makeatother

\newcommand{\MyMakeDupTheoMacros}[7]{
\MyMakeTheoMacros{#1}{#2}{#3}
\newreptheorem{#1}{#6}
\newcommand{#4}[3]{
\newcommand{##2}{##3}
\begin{#1}\label{##1}~##2\end{#1}}
\newcommand{#5}[4]{
\newcommand{##2}{##4}
\begin{#1}{\e{##3}}\label{##1}~##2\end{#1}}
\newcommand{#7}[2]{\begin{rep#1}{##1}~##2 \end{rep#1}}
}

{}
\newcommand{\MyMakeRefMacros}[3]{\newcommand{#1}[2][]
{\ifthenelse{\equal{}{##1}}{#2~\ref{##2}}{#3~\ref{##1} and~\ref{##2}}}}

\newcommand{\MyMakeEqRefMacros}[3]{\newcommand{#1}[2][]
{\ifthenelse{\equal{}{##1}}{#2~\eqref{##2}}{#3~\eqref{##1} and~\eqref{##2}}}}
{}

{}  %
\newcommand{\bibentry}[8]{
{}\bibitem[\nospell{#8}]{#1} {\textup #3}.{}
\ifthenelse{\equal{}{#6}}
{\newblock \textrm{#4.} \newblock {\em #5}, #7....}
{\newblock \textrm{#4.} \newblock {\em #5, #6}, #7.}
}

{} %

\MyMakeTheoMacros{fact}{\fct}{\nfct}

\MyMakeRefMacros{\fctref}{\dgFact}{\dgFacts}
\MyMakeRefMacros{\Fctref}{\dgCapFact}{\dgCapFacts}

\MyMakeTheoMacros{question}{\quest}{\nquest}

\MyMakeRefMacros{\questref}{\dgQuestion}{\dgQuestions}
\MyMakeRefMacros{\Questref}{\dgCapQuestion}{\dgCapQuestions}

{} %
\MyMakeDupTheoMacros{lemma}
{\lem}{\nlem}{\lemdup}{\nlemdup}{\dgLemma}{\lemrep}
{}

\MyMakeRefMacros{\lemref}{\dgLemma}{\dgLemmas}
\MyMakeRefMacros{\Lemref}{\dgCapLemma}{\dgCapLemmas}

\newcommand{\fakelemref}[1]{\dgLemma~{#1}}

\MyMakeDupTheoMacros{corollary}
{\crl}{\ncrl}{\crldup}{\ncrldup}{\dgCorollary}{\crlrep}

\MyMakeRefMacros{\crlref}{\dgCorollary}{\dgCorollaries}
\MyMakeRefMacros{\Crlref}{\dgCapCorollary}{\dgCapCorollaries}

\MyMakeTheoMacros{proposition}{\prp}{\nprp}

{}
\newtheorem*{prp*}{\e{\dgCapProposition}}

{}

\MyMakeRefMacros{\prpref}{\dgProposition}{\dgPropositions}
\MyMakeRefMacros{\Prpref}{\dgCapProposition}{\dgCapPropositions}

\MyMakeDupTheoMacros{my_claim}
{\clm}{\nclm}{\clmdup}{\nclmdup}{\dgClaim}{\clmrep}

\MyMakeRefMacros{\clmref}{\dgClaim}{\dgClaims}
\MyMakeRefMacros{\Clmref}{\dgCapClaim}{\dgCapClaims}

\MyMakeDupTheoMacros{theorem}
{\theo}{\ntheo}{\theodup}{\ntheodup}{\dgTheorem}{\theorep}

\MyMakeRefMacros{\theoref}{\dgTheorem}{\dgTheorems}
\MyMakeRefMacros{\Theoref}{\dgCapTheorem}{\dgCapTheorems}

\MyMakeTheoMacros{postulate}{\postu}{\npostu}

\MyMakeRefMacros{\posturef}{\dgPostulate}{\dgPostulates}
\MyMakeRefMacros{\Posturef}{\dgCapPostulate}{\dgCapPostulates}

\MyMakeTheoMacros{definition}{\defi}{\ndefi}

\MyMakeRefMacros{\defiref}{\dgDefinition}{\dgDefinitions}
\MyMakeRefMacros{\Defiref}{\dgCapDefinition}{\dgCapDefinitions}

\MyMakeTheoMacros{problem}{\prob}{\nprob}

\MyMakeRefMacros{\probref}{\dgProblem}{\dgProblems}
\MyMakeRefMacros{\Probref}{\dgCapProblem}{\dgCapProblems}

\MyMakeTheoMacros{myremark}{\rem}{\nrem}

\MyMakeRefMacros{\remref}{\dgRemark}{\dgRemarks}
\MyMakeRefMacros{\Remref}{\dgCapRemark}{\dgCapRemarks}

\MyMakeTheoMacros{conjecture}{\conj}{\nconj}

\MyMakeRefMacros{\conjref}{\dgConjecture}{\dgConjectures}
\MyMakeRefMacros{\Conjref}{\dgCapConjecture}{\dgCapConjectures}

\renewcommand{\qedsymbol}{$\blacksquare$}

\newcommand{\prfstart}[1][]{\ifthenelse{\equal{}{#1}}%
{\begin{proof}\renewcommand{\qedsymbol}{$\blacksquare$}}%
{\begin{proof}[\dgProofOf\ #1]%
\renewcommand{\qedsymbol}{$\blacksquare_{\mbox{\it{\scriptsize{#1}}}}$}}%
}
\newcommand{\prfend}[1][*]{%
\ifthenelse{\equal{}{#1}}{\renewcommand{\qedsymbol}{$\blacksquare$}}{}%
\ifthenelse{\equal{*}{#1}}{}%
{\renewcommand{\qedsymbol}{$\blacksquare_{\mbox{\it{\scriptsize{#1}}}}$}}%
\end{proof}\renewcommand{\qedsymbol}{$\blacksquare$}%
}

\newcommand{\sect}[2][]{
\ifthenelse{\equal{*}{#2}}
{\section*}
{\ifthenelse{\equal{}{#1}}
{\section{#2}}
{\section{#2}\label{#1}}
}
}

\newcommand{\ssect}[2][]{
\ifthenelse{\equal{*}{#2}}
{\subsection*}
{\ifthenelse{\equal{}{#1}}
{\subsection{#2}}
{\subsection{#2}\label{#1}}
}
}

\newcommand{\sssect}[2][]{
\ifthenelse{\equal{*}{#2}}
{\subsubsection*}
{\ifthenelse{\equal{}{#1}}
{\subsubsection{#2}}
{\subsubsection{#2}\label{#1}}
}
}

\newcommand{\para}[2][]{\ifthenelse{\equal{}{#1}}
{\paragraph{#2}}
{\paragraph{#2}\label{#1}}}

\MyMakeRefMacros{\chref}{\dgChapter}{\dgChapters}
\MyMakeRefMacros{\Chref}{\dgCapChapter}{\dgCapChapters}

\MyMakeRefMacros{\sref}{\dgSection}{\dgSections}
\MyMakeRefMacros{\Sref}{\dgCapSection}{\dgCapSections}

\MyMakeRefMacros{\ssref}{\dgSubsection}{\dgSubsections}
\MyMakeRefMacros{\Ssref}{\dgCapSubsection}{\dgCapSubsections}

\MyMakeRefMacros{\sssref}{\dgSubsection}{\dgSubsections}
\MyMakeRefMacros{\Sssref}{\dgCapSubsection}{\dgCapSubsections}

\MyMakeRefMacros{\figref}{\dgFigure}{\dgFigures}
\MyMakeRefMacros{\Figref}{\dgCapFigure}{\dgCapFigures}

{}
\newcommand{\MyChangeMathMargins}{%
\setlength{\abovedisplayskip}{\abovedisplayshortskip + 2pt}%
\setlength{\belowdisplayskip}{\belowdisplayshortskip + 1pt}%
}
{}

\newcommand{\IfMathMode}[2]{\ifmmode{#1}\else{#2}\fi}

\newcommand{\Ensuremath}{\ensuremath}

\newcommand{\fbr}[1]{\IfMathMode%
{#1}{$#1$}}                     %

\newcommand{\fnbr}[1]{\mbox{\fbr{#1}}}  %

\newcommand{\fla}[2][*]{\ifthenelse{\equal{}{#1}}{\fbr{#2}}{\fnbr{#2}}}
\newcommand{\f}{\fla}

\newcommand{\malabel}[1]{\addtocounter{equation}{1}\tag{\theequation}\label{#1}}
\newcommand{\mal}[2][]{\MyChangeMathMargins%
\ifthenelse{\equal{}{#1}}%
{\begin{align*} #2 \end{align*}}%
{\ifthenelse{\equal{P}{#1}}%
{\allowdisplaybreaks\begin{align*} #2%
\end{align*}\interdisplaylinepenalty=10000}%
{\begin{align*} \malabel{#1} #2 \end{align*}}%
}%
}

\newcommand{\m}{\mal}

\newcommand{\mac}{\substack}

\MyMakeEqRefMacros{\equref}{\dgEquation}{\dgEquations}
\MyMakeEqRefMacros{\Equref}{\dgCapEquation}{\dgCapEquations}

\MyMakeEqRefMacros{\expref}{\dgExpression}{\dgExpressions}
\MyMakeEqRefMacros{\Expref}{\dgCapExpression}{\dgCapExpressions}

\MyMakeEqRefMacros{\inequref}{\dgInequality}{\dgInequalities}
\MyMakeEqRefMacros{\Inequref}{\dgCapInequality}{\dgCapInequalities}

\newcommand{\bref}[1]{(\ref{#1})}

\newcommand{\twocase}[4]%
{\begin{cases} #1 &\txt{#2}\\ #3 &\txt{#4}\end{cases}}

\newcommand{\thrcase}[6]%
{\begin{cases} #1 &\txt{#2}\\ #3 &\txt{#4}\\ #5 &\txt{#6}\end{cases}}

\newcommand{\fourcase}[8]%
{\begin{cases} #1 &\txt{#2}\\ #3 &\txt{#4}\\ #5 &\txt{#6}\\ #7 &\txt{#8}\end{cases}}

\newcommand{\lf}{\mathopen{}\mathclose\bgroup\left}
\newcommand{\rt}{\aftergroup\egroup\right}

\providecommand{\middle}{\big}
\newcommand{\md}{\middle}

\newcommand{\chs}{\genfrac(){0cm}{}}  %

\newmatop{\poly}{poly}
\newmatop[disc]{\disc}{disc_{#1}}
\newmatop{\Pow}{Pow}
\newmatop{\supp}{supp}   %

\makeatletter
\def\moverlay{\mathpalette\mov@rlay}
\def\mov@rlay#1#2{\leavevmode\vtop{%
\baselineskip\z@skip \lineskiplimit-\maxdimen
\ialign{\hfil$\m@th#1##$\hfil\cr#2\crcr}}}
\newcommand{\charfusion}[3][\mathord]{
#1{\ifx#1\mathop\vphantom{#2}\fi
\mathpalette\mov@rlay{#2\cr#3}
}
\ifx#1\mathop\expandafter\displaylimits\fi}
\makeatother

\providecommand{\cupdot}{\charfusion[\mathbin]{\cup}{\cdot}}

\newcommand{\h}[2][]{\ifthenelse{\equal{}{#2}}%
{\mathop H_{#1}}%
{\mathop H_{#1}{\l({#2}\r)}}}
\newcommand{\hh}[3][]{\mathop H_{#1}%
{\l({#2}\vphantom{|_1^1}\md|\vphantom{|_1^1}{#3}\r)}}

\newcommand{\KL}[2]{d_{KL}\lf({#1}\md\|\vphantom{|_1^1}{#2}\rt)}

\newcommand{\I}[3][]{\ifthenelse{\equal{}{#1}}%
{\pmb{I}\lf[{#2}\vphantom{|_1^1}:\vphantom{|_1^1}{#3}\rt]}%
{\mathop{\pmb{I}}_{#1}\lf[{#2}\vphantom{|_1^1}:\vphantom{|_1^1}{#3}\rt]}%
}

\newcommand{\Ii}[4][]{\ifthenelse{\equal{}{#1}}%
{\pmb{I}\lf[{#2}\vphantom{|_1^1}:\vphantom{|_1^1}{#3}\md|{#4}\rt]}%
{\mathop{\pmb{I}}_{#1}\lf[{#2}\vphantom{|_1^1}:\vphantom{|_1^1}{#3}\md|{#4}\rt]}%
}

\providecommand{\E}[2][]{\mathop{\pmb{E}}_{#1}\lf[{#2}\rt]}
\newcommand{\Ee}[3][]
{\mathop{\pmb{E}}_{#1}\lf[{#2}\vphantom{|_1^1}\md|\vphantom{|_1^1}{#3}\rt]}

\newcommand{\PR}[2][]{\mathop{\pmb{Pr}}_{#1}\lf[{#2}\rt]}
\newcommand{\PRr}[3][]{\mathop{\pmb{Pr}}_{#1}\lf[{#2}\vphantom{|_1^1}\md|\vphantom{|_1^1}{#3}\rt]}

\renewcommand{\U}[1][]{\ifthenelse{\equal{}{#1}}%
{{\cal U}}%
{{\cal U}_{#1}}}

\providemat{\QQ}{\mathbb{Q}}
\providemat{\NN}{\mathbb{N}}
\providemat{\CC}{\mathbb{C}}
\providemat{\RR}{\mathbb{R}}
\providemat{\ZZ}{\mathbb{Z}}

\newcommand{\wtl}{\widetilde}
\newcommand{\wbr}{\overline}   %

\newcommand{\pss}[1][]{\nospell{\ifthenelse{\equal{}{#1}}%
{\txt{'s}}%
{\fla{#1\txt{'s}}}}}

\newcommand{\pl}[1][]{\nospell{\ifthenelse{\equal{}{#1}}%
{\mskip-6mu\stackrel{\text-}{}\mskip-4mu\txt{s}}%
{\fla{#1\mskip-6mu\stackrel{\text-}{}\mskip-4mu\txt{s}}}}}

\newcommand{\ord}[1][]{\nospell{\ifthenelse{\equal{}{#1}}%
{\txt{'th}}%
{\ifthenelse{\equal{1}{#1}}{$1\txt{'st}$}{\ifthenelse{\equal{2}{#1}}{$2\txt{'nd}$}{\ifthenelse{\equal{3}{#1}}{$3\txt{'rd}$}{\fla{#1\txt{'th}}}}}}}}

\newcommand{\fr}[3][*]{%
\ifthenelse{\equal{*}{#1}}%
{\frac{#2}{#3}}{}%
\ifthenelse{\equal{/}{#1}}%
{\nicefrac{#2}{#3}}{}%
\ifthenelse{\equal{}{#1}}%
{\lf.#2\md/#3\rt.}{}%
\ifthenelse{\equal{p_}{#1}}%
{\lf.\lf(#2\rt)\md/#3\rt.}{}%
\ifthenelse{\equal{_p}{#1}}%
{\lf.#2\md/\lf(#3\rt)\rt.}{}%
\ifthenelse{\equal{pp}{#1}}%
{\lf.\lf(#2\rt)\md/\lf(#3\rt)\rt.}{}%
}

\newcommand{\dr}{\nicefrac}

\newcommand{\sq}{\sqrt}

\newcommand{\set}[2][]{\ifthenelse{\equal{}{#1}}%
{\Ensuremath{\lf\{#2\rt\}}}%
{\Ensuremath{\lf\{#2\vphantom{|_1^1}\md|\vphantom{|_1^1}#1\rt\}}}}

\newcommand{\sett}[2]{\Ensuremath{\lf\{#1\vphantom{|_1^1}\md|\vphantom{|_1^1}#2\rt\}}}

\newcommand{\Log}[2][]{\ifthenelse{\equal{}{#1}}%
{\log\lf(#2\rt)}%
{\log_{#1}\lf(#2\rt)}%
}

\newcommand{\Minn}[3][]{\ifthenelse{\equal{}{#1}}%
{\Ensuremath{\min_{#2}\lf\{#3\rt\}}}%
{\Ensuremath{\min_{#2}\lf\{#3\vphantom{|_1^1}\md|\vphantom{|_1^1}#1\rt\}}}}

\newcommand{\Maxx}[3][]{\ifthenelse{\equal{}{#1}}%
{\Ensuremath{\max_{#2}\lf\{#3\rt\}}}%
{\Ensuremath{\max_{#2}\lf\{#3\vphantom{|_1^1}\md|\vphantom{|_1^1}#1\rt\}}}}

\newfunction{\asO}{O}
\newfunction{\aso}{o}
\newfunction{\asOm}{\Omega}
\newfunction{\astOm}{\tilde \Omega}
\newfunction{\asT}{\Theta}

\providecommand{\ket}[1]{\Ensuremath{\lf|#1\rra}}

\newcommand{\bket}[3][]{\ifthenelse{\equal{}{#1}}%
{\Ensuremath{\lla #2\md|#3\rra}}%
{\Ensuremath{\lla #2\md|#1\md|#3\rra}}
}

\providecommand{\ip}[2]{\Ensuremath{\lla #1\,,\,#2\rra}}

\newcommand{\sz}[2][]{\ifthenelse{\equal{}{#1}}%
{\Ensuremath{\lf|#2\rt|}}%
{\Ensuremath{\lf|#2\rt|_{#1}}}}
\providecommand{\norm}[2][]{\ifthenelse{\equal{}{#1}}%
{\Ensuremath{\lf\|#2\rt\|}}%
{\Ensuremath{\lf\|#2\rt\|_{#1}}}}

\newcommand{\bm}{\pmb} %

\newcommand{\txt}[1]{\textrm{#1}}  %

\newcommand{\Cl}{\mathcal}  %

\DeclareMathAlphabet{\mathlowcal}{OT1}{pzc}{m}{it}
\newcommand{\Cll}{\mathlowcal}

\newident[]{\DII}{\mathcal D_{#1}^{\parallel}}
\newident[]{\R}{\mathcal R_{#1}}
\newident[]{\RI}{\mathcal R_{#1}^{1}}
\newident[]{\RII}{\mathcal R_{#1}^{\parallel}}
\newident[]{\RIIp}{\mathcal R_{#1}^{\parallel,pub}}
\newident[]{\RIIe}{\mathcal R_{#1}^{\parallel,ent}}
\newident[]{\Q}{\mathcal Q_{#1}}
\newident[]{\QI}{\mathcal Q_{#1}^{1}}
\newident[]{\QII}{\mathcal Q_{#1}^{\parallel}}
\newident[]{\QIIp}{\mathcal Q_{#1}^{\parallel,pub}}

\newidenT{\Pp}{P}
\newidenT[BPP]{\BPP}{BPP_{#1}}
\newidenT{\ZPP}{ZPP}
\newidenT{\SBP}{SBP}
\newidenT{\coSBP}{coSBP}
\newidenT[RP]{\RP}{RP_{#1}}
\newidenT{\PP}{PP}
\newidenT{\UPP}{UPP}
\newidenT{\coRP}{coRP}
\newidenT{\BQP}{BQP}
\newidenT{\NP}{NP}
\newidenT{\coNP}{coNP}
\newidenT{\AM}{AM}
\newidenT[MA]{\MA}{MA_{#1}}
\newidenT[QMA]{\QMA}{QMA_{#1}}
\newidenT[ZAM]{\ZAM}{ZAM_{#1}}
\newidenT[UAM]{\UAM}{UAM_{#1}}
\newidenT{\PH}{PH}
\newidenT{\PSPACE}{PSPACE}
\newidenT{\EXP}{EXP}
\newidenT{\NEXP}{NEXP}

\newidenT{\DNF}{DNF}
\newidenT{\Eq}{Eq}
\newidenT{\Disj}{Disj}
\newidenT{\IP}{IP}
\newident{\tEq}{\wtl{Eq}}

\newmat{\mset}{\smin\set}

\newcommand{\lla}{\lf\langle}
\newcommand{\rra}{\rt\rangle}

\newcommand{\nin}{\not\in}  %

\newcommand{\Then}{\Longrightarrow}

\newcommand{\dt}{\cdot}
\newcommand{\tm}{\cdot}
\newcommand{\xor}{\oplus}
\newcommand{\sdif}{\vartriangle}  %
\newcommand{\sbseq}{\subseteq}
\newcommand{\sbs}{\subset}
\newcommand{\smin}{\setminus}
\newcommand{\eps}{\varepsilon}
\newcommand{\deq}{\stackrel{\textrm{def}}{=}}

\newcommand{\unin}{\mathrel{\subset\mkern-13.1mu\sim}}  %

\newcommand{\ds}[1][]
{\ifthenelse{\equal{}{#1}}{\allowbreak\dots}{#1\allowbreak\dots#1}}
\newmat{\dc}{\ds[,]}
\newmat{\dtimes}{\ds[\times]}

\mydef{\01}{\set{0,1}}

\mydef{\l(}{\lf(}
\mydef{\r)}{\rt)}

\mydef{\=}{\equiv}
\mydef{\+}{\oplus}

\mathchardef\myhyphen="2D
\mydef{\-}{\myhyphen}

\newcommand{\abstr}[1]{\begin{abstract} #1 \end{abstract}}

\newcommand{\cent}[1]{\begin{center} #1 \end{center}}

\newcommand{\red}{\textcolor{Red}}

{} %
\newcommand{\itemi}[2][\MyUniPat]{\ifthenelse{\equal{\MyUniPat}{#1}}%
{\begin{itemize}[noitemsep,topsep=3pt] #2 \end{itemize}}%
{\begin{itemize}[#1] #2 \end{itemize}}}
\newcommand{\enum}[2][\MyUniPat]{\ifthenelse{\equal{\MyUniPat}{#1}}%
{\begin{enumerate}[noitemsep,topsep=3pt] #2 \end{enumerate}}%
{\begin{enumerate}[#1] #2 \end{enumerate}}}

\newcommand{\itstart}[1][\MyUniPat]{\ifthenelse{\equal{\MyUniPat}{#1}}%
{\begin{itemize}[noitemsep,topsep=3pt]}%
{\begin{itemize}[#1]}%
}

{}  %

\newcommand{\itend}{\end{itemize}}

\protected \def \dg #1{%
\textcolor{Red}
{
{\normalmarginpar\marginnote{\bl{DG's comment}}}
{\reversemarginpar\marginnote{\bl{DG's comment}}\\}
\IfMathMode{
~~~\txt{#1}~
}{
~\\~~~#1~\\
{\normalmarginpar\marginnote{\bl{\ul{------}}}}
{\reversemarginpar\marginnote{\bl{\ul{------}}}\\}
}
}
\ClassWarning{My Macros}{#1}
}

\newcommand{\fn}[2][]{%
\IfMathMode{}{}%
\ifthenelse{\equal{}{#1}}%
{\footnote{#2}}%
{\footnote{\label{#1}#2}}%
}



\DeclareTextFontCommand{\bemph}{\bfseries}
\DeclareTextFontCommand{\ibemph}{\bfseries\em}
{} %
\newcommand{\e}{\emph}

{}  %

\newcommand{\bl}[1]{{\bf #1}} %

\newcommand{\il}[1]{{\it #1}} %

\providecommand{\ul}[1]{\underline{#1}} %

\newcommand{\tb}{\quad}
\newcommand{\tbb}{\qquad}
\newcommand{\tbbb}{\qquad\qquad}

\makeatother%

{}

{}

\makeatother%

\newident{\Sh}{Shape}

\newident{\Equ}{\Eq_u}
\newident{\EquT}{\Eq_{u,T}}
\newident{\EqT}{\Eq_T}
\newident{\tEqT}{\wtl{Eq}_T}
\newident{\ctEqT}{\wtl{cEq}_T}

\newident{\mEqu}{\mu_{\Equ}}
\newident{\mEquT}{\mu_{\EquT}}
\newident{\mtEqT}{\mu_{\tEqT}}
\newident{\mctEqT}{\mu_{\ctEqT}}

\title{Quantum versus classical simultaneity\\
in communication complexity}

\newcommand{\instDG}{Institute of Mathematics, Czech Academy of Sciences, \v Zitna 25, Praha 1, Czech Republic.}

\newcommand{\thanksDG}{Partially funded by the grant 19-27871X of GA \v CR.
Part of this work was done while visiting the Centre for Quantum Technologies at the National University of Singapore, and was partially supported by the Singapore National Research Foundation, the Prime Minister's Office and the Ministry of Education under the Research Centres of Excellence programme under grant R 710-000-012-135.}

{}
\author{Dmitry Gavinsky\thanks{\instDG\ \thanksDG}
}
{}

\begin{document}

\maketitle

\thispagestyle{empty}

\abstr{
This work addresses two problems in the context of two-party communication complexity of functions.
First, it concludes the line of research which can be viewed as demonstrating qualitative advantage of quantum communication in the three most common communication ``layouts'':\ two-way interactive communication; one-way communication; simultaneous message passing (SMP).
We demonstrate a functional problem $\ctEqT$, whose communication complexity is \asO{(\log n)^2} in the quantum version of SMP and \astOm{\sq n} in the classical (randomised) version of SMP.

Second, this work contributes to understanding the power of the weakest commonly studied regime of quantum communication -- SMP with quantum messages and \e{without shared randomness} (the latter restriction can be viewed as a somewhat artificial way of making the quantum model ``as weak as possible'').
Our function $\ctEqT$ has an efficient solution in this regime as well, which means that even lacking shared randomness, quantum SMP can be exponentially stronger than its classical counterpart with shared randomness.

}

\sect{Introduction}

Communication complexity is among the most interesting computational realms so far:
Being one of the strongest where we can establish non-trivial (often tight) hardness statements -- \e{lower bounds}; at the same time, it is one of the weakest that is capable to ``accommodate'' rather involved algorithms -- \e{protocols}.
As of today, communication complexity is one of the very few computational scenarios where both upper and (non-speculative) lower bounds play central roles in the research.

We address two questions, related to the most basic communication complexity setting -- the regime of \e{two parties}, solving a \e{functional problem}.

\para{Two-way, one-way and SMP.}
The three most commonly studied bipartite communication ``layouts'' are:\ \e{two-way (interactive) communication}, \e{one-way communication} and \e{simultaneous message passing (SMP)}.
These models involve two players, \e{Alice} and \e{Bob}, who receive one ``portion'' of the input each:\ Alice gets $X$ and Bob gets $Y$ (which we view as random variables).
Their goal is to use the allowed type of communication (as determined by the ``layout'', see below) in order to compute the value of $f(X,Y)$, where $f$ is a two-argument function defining the computational problem that the players have to solve.

\itemi{
\item In the model of two-way communication the players can exchange messages, until one of them outputs the answer.
\item In the one-way model Alice can send one message to Bob, who then produces the answer, based on this message and his portion of the input.
\item In the model of simultaneous message passing both Alice and Bob send one message each to the third participant -- the \e{referee} -- who has to produce the answer, based on these two messages only (unlike the players, the referee doesn't directly receive any portion of the input).
}

In all three regimes we say that a communication protocol \e{computes} a Boolean function $f$ if for every pair $(x,y)$ from the support of $f$, when the players receive $(X,Y)=(x,y)$, they output $f(x,y)$ with probability at least $\dr23$.
The participants are ``all powerful'' in terms of their local computational abilities, and the only resource considered for determining the \e{cost} of a protocol is the ``amount of communication'' that it consumes.

\itemi{
\item When the communication model is \e{randomised}, the participants can send (classical) bits, the correctness condition must hold with respect to the random choices made by them and the complexity of a protocol is the (maximum) total number of bits sent during its execution.
\item When the model is \e{quantum}, the participants can send qubits and perform arbitrary quantum measurements, the correctness condition must hold with respect to these quantum operations and the complexity of a protocol is the (maximum) total number of qubits sent during its execution.
}

It is known (and easy to see) that for virtually any type of communication ``primitive'' (i.e., classical randomised; classical deterministic; quantum; ...), the two-way layout is the most powerful, one-way is intermediate and SMP is the weakest.

Demonstrating advantage of quantum over classical communication in a weaker regime (say, one-way) could -- in principle -- turn out to be either less or more challenging than in a stronger one (say, interactive):
While in the latter case one would have to prove a \e{stronger} lower bound, at the same time the communication problem being used for the separation would likely be \e{harder}, and therefore easier to prove a lower bound for.

The history of research seems to suggest that separating models on the ``lower levels'' -- namely, one-way communication, and even more so SMP -- is more challenging than under the stronger setting of interactive communication.
In 1999 Raz~\cite{R99_Ex} demonstrated a \e{function} that had an efficient\fn
{
We call \e{efficient} communication protocols whose complexity is poly-logarithmic in the input length.
}
\e{quantum two-way protocol}, but no efficient \e{classical two-way protocol}.
In 2004 Bar-Yossef, Jayram and Kerenidis~\cite{BJK04_Exp} demonstrated a \e{relation} that had an efficient \e{quantum one-way protocol}, but no efficient \e{classical one-way protocol}.
Note that the original separation from~\cite{R99_Ex} was demonstrated via a functional problem; on the other hand, the result of~\cite{BJK04_Exp} used a relation -- a more general class of problems and a \e{stronger} model-separating tool.\fn
{
There are known cases where a quantum communication model can be separated from a classical one via a relation, but a functional separation is provably impossible (see~\cite{A04_Lim,GRW08_Sim}).
In particular,~\cite{GRW08_Sim} showed that the class of \e{functional} problems, efficiently computable in ``quantum-classical SMP'' -- the regime where Alice could send a quantum message but Bob was classical (or vice versa) -- was equal to the corresponding class of the ``fully classical'' SMP regime; on the other hand, a \e{relational} separation between these two models followed from~\cite{BJK04_Exp}.
As in this work we are only concerned with super-polynomial separations via functional problems, for us the model of SMP with both players being quantum is the weakest (non-trivial) regime of quantum communication.
}
In the same work it has been asked whether it was possible to demonstrate similar qualitative advantage of quantum one-way communication via a functional problem, which was answered affirmatively in 2008 in a joint work with Kempe, Kerenidis, Raz and de Wolf~\cite{GKKRW08_Ex}.

The work~\cite{BJK04_Exp} has also demonstrated a \e{relation} that had an efficient \e{quantum SMP protocol}, but no efficient \e{classical SMP protocol}, and -- similarly to the one-way case -- it has been left open whether there existed a functional problem, easy for quantum and hard for classical SMP.

In the meantime, separations ``against classical two-way'' have been strengthened in a sequence of works~\cite{G08_Cla,KR11_Qua,G16_En} that subsumed earlier separations:\ e.g., in 2010 Klartag and Regev demonstrated a \e{function} with an efficient \e{quantum one-way protocol}, but no efficient \e{classical two-way protocol}.
On the other hand, it has remained open till now whether a function could witness quantum superiority in the case of SMP.\fn[fn_G16]
{
The result in~\cite{G16_En} implied existence of a function, hard for classical SMP (and even for classical two-way protocols), but easy for the model of \e{quantum SMP with shared entanglement} -- a significantly strengthened version of quantum SMP, where the players could share an arbitrary (input-independent) quantum state of finite dimension.
}

This work presents a functional problem $\ctEqT$, whose communication complexity is \asO{(\log n)^2} in the quantum version of SMP and \astOm{\sq n} in the classical (randomised) version of SMP.

\para{Weakening the weak: SMP without shared randomness.}
The second aspect of this work is related to understanding the power of, arguably, the weakest commonly studied regime of quantum communication -- SMP with quantum messages and \e{without shared randomness}.

We will write \QII\ and \RII\ for, respectively, the quantum and the classical version of the model SMP without shared randomness.
To denote the corresponding standard counterparts -- those equipped with (unlimited) shared randomness -- we will write, respectively, \QIIp\ and \RIIp.
For any model $\Cl M$ and a problem $\Cl P$, we will write $\Cl M(\Cl P)$ to denote the complexity of $\Cl P$ in $\Cl M$.

Both \QII\ and \RII\ (i.e., the versions lacking shared randomness) can be viewed as ``purposely weakened'', somewhat artificial versions of SMP -- as opposed to the standard \QIIp\ and \RIIp.\fn
{
Note that in the context of ``Two-way, one-way and SMP'' we only referred to the ``natural'' models \QIIp\ and \RIIp.
}
The families of efficiently-computable tasks in \QII\ and in \RII\ are not closed with respect to mixed strategies\fn
{
\RIIp\ -- the ``unrestricted'' randomised SMP -- can be defined as the ``closure'' of \RII\ with respect to mixed strategies, and similarly for \QIIp\ and \QII.
}, and the usual minimax principle does not hold for these models:\ for example, the \e{equality function (\Eq)} has \RII-complexity \asO1 over any fixed input distribution, but its worst-case \RII-complexity is \asOm{\sqrt n}, due to~\cite{NS96_Pub}.

Von Neumann, who proved the minimax principle for the case of 2-player zero-sum games with mixed strategies in 1928, later remarked:\ ``\e{As far as I can see, there could be no theory of games [...] without that theorem.}''
The question of determining the complexity of a given communication problem can be phrased in the language of 2-player zero-sum games, and the case of SMP without shared randomness is probably the only commonly studied one that goes ``without that theorem''.
Although we have seen some non-trivial results both in \QII\ and in \RII, these models still lack the aesthetic appeal and the cognitive depth of those obeying the minimax principle.

So, the model of SMP with quantum messages and without shared randomness (\QII) indeed can be viewed as the weakest commonly studied quantum model in communication complexity.
Prior to this work, \QII\ was known to be stronger than \RII:\ in 2001 Buhrman, Cleve, Watrous and de Wolf~\cite{BCWW01_Qua} demonstrated that there existed a \QII-protocol for the function \Eq\ of complexity \asO{\log n}; as we already mentioned, it had been known that $\RII(\Eq)\in\asOm{\sqrt n}$.
Till now it has remained open whether \QII\ was capable to do more than that -- in particular, to solve efficiently any problem that was hard for the ``natural closure'' of \RII, namely \RIIp.

We show that the main communication problem studied in this work -- the function $\ctEqT$ -- has an efficient protocol in \QII\ as well.
Due to the same lower bound of \astOm{\sq n} on its \RIIp-complexity, this demonstrates exponential advantage of \QII\ over \RIIp\ in solving a functional problem.

One obvious question that remains open is whether there is a bipartite communication problem -- even a relational one -- that admits an efficient solution in \QII, though not in \R.\fn
{
As a black-box statement, demonstrating a \e{functional} problem with those properties (whose existence one may question:\ even a \e{relational} separation like that is not presently known) would subsume the current work, as well as~\cite{G16_En}.
On the other hand, here we demonstrate a lower bound in \RIIp\ for a functional problem that is, intuitively, very close to being within the reach of this model (as witnessed, in particular, by the fact that the problem is easy for \QII).
The aesthetic appeal of the quest of finding an appropriate fine-tuned analytic approach has been the author's main motivation for addressing this question.
}
Further historical background and some open questions can be found in \sref{s_land}.

\para{Why this is interesting technically.}
As a part of this work, \RIIp-hardness is argued for a communication problem, which is easy for virtually any model stronger than \RIIp.
Therefore the argument has to be tuned rather accurately in order to distinguish between \RIIp\ and some other models of communication that are ``just slightly stronger'' (like \RI).

On the other hand, the complexity of the analysed communication task must also be tuned, as it has to be easy for \QII\ and hard for \RIIp, which is ``just slightly weaker'' (sometimes even incomparable\footnotemark).
\footnotetext
{
There are known examples, where \RIIp\ is exponentially stronger than \QII\ for relational problems, see~\cite{GKRW09_Bo}.
}
In particular we cannot use a problem with \e{worst-case hardness in spite of average-case easiness} (like \Eq), as \RIIp\ allows for mixed strategies.

It may be for these reasons that this work is built around several ad hoc ideas.\fn
{
Let us remark that technically this work is very different from~\cite{G16_En} -- except for the definitions of the core communication tasks that are considered, which share a few obvious structural similarities (e.g., both the problems are naturally viewed as ``distant derivatives'' from the equality problem).
We do not know whether \Sh\ -- the core task of~\cite{G16_En} -- admits an efficient \QII-, or even \QIIp-protocol (and conjecture that it doesn't); on the other hand, the core task of the current work -- \ctEqT\ -- is trivial not only for \R, but even for \RI\ (see Sect.~\ref{ss_tow_low}).
}
Some of them will be informally discussed in \sref{s_sum}.

\sect[s_prelim]{Preliminaries}

For $x\in\01^n$ and $i\in{[n]}=\set{1\dc n}$, we will write $x_i$ or $x(i)$ to address the \ord[i] bit of $x$ (preferring ``$x_i$'' unless it may cause ambiguity).
Similarly, for $S\sbseq{[n]}$, let both $x_S$ and $x(S)$ denote the $\sz S$-bit string, consisting of (naturally-ordered) bits of $x$, whose indices are in $S$.
For a set (or a family) $A$, we will write $A|_i$ and $A|_S$ to address, respectively, $\sett{x_i}{x\in A}$ and $\sett{x_S}{x\in A}$.
We will use similar notation in all cases when $x$ can be viewed naturally as an element of $\Cl X_1\dtimes\Cl X_n$.

For $x,\,y\in\01^n$, let $\sz x$ denote the Hamming weight of $x$ and $x\xor y$ denote the bit-wise XOR operation.

For a (discrete) set $A$ and $k\in\NN$, we denote by $\Pow(A)$ the set of \pss[A] subsets and by $\chs Ak$ the set $\sett{a\in\Pow(A)}{\sz a=k}$.
We write ``$A\sdif B$'' to denote the symmetric difference between the two sets and ``$A\cupdot B$'' to denote the union when $A$ and $B$ are disjoint (i.e., writing ``$A\cupdot B$'' implies that $A\cap B=\emptyset$).

We write $\U[A]$ to denote the uniform distribution over the elements of $A$.
Sometimes (e.g., in subscripts) we will write ``$\unin A$'' instead of ``$\sim\U[A]$''.
We will sometimes emphasise that a distribution on $\01^{2n}$ is ``viewed as bipartite'' (i.e., assumed to be the joint distribution of two random variables, containing $n$ bits each) by addressing it as a \e{distribution on $\01^{n+n}$}; similarly, we will write ``$(X,Y)\in\01^{n+n}$'', etc.

For (discrete) distributions $\mu_1$ and $\mu_2$, their \e{relative entropy} is
\m{
\KL{\mu_1}{\mu_2}
\deq \sum_{x\in\supp(\mu_1)\cup\supp(\mu_2)}
\mu_1(x) \tm \log\l(\fr{\mu_1(x)}{\mu_2(x)}\r)
,}
where the logarithm is base-$2$.
It follows readily from the strict concavity of $\log$ that
\m{
\KL{\mu_1}{\mu_2} \ge 0 
,}
where the equality holds if and only if $\mu_1\=\mu_2$.

We will use the Chernoff bound in the following form.

\nfct[f_Cher]{tail-estimating inequalities}{For $n\in\NN$, let $\bar X=(X_1\dc X_n)\sim\mu$ be mutually independent random variables, satisfying $\E[\mu]{X_i}\=p\in[0,1]$.
Then for any $\alpha\in\asOm1$:
\m{
\PR[\mu]{\sum_{i=1}^n X_i \ge (p+\alpha)\tm n},\:
\PR[\mu]{\sum_{i=1}^n X_i \le (p-\alpha)\tm n} \, \in \, 2^{-\asOm n}
.}

Let $\mu'$ be any distribution, satisfying $\norm[1]{\mu-\mu'}\le\beta$, then
\m{
\PR[\mu']{\sum_{i=1}^n X_i \ge (p+\alpha)\tm n},\:
\PR[\mu']{\sum_{i=1}^n X_i \le (p-\alpha)\tm n} \, \in \, 2^{-\asOm n} + \fr\beta2
.}
}

Let $S_n$ denote the group of permutations of $[n]$, and let $\sigma_i\in S_n$ be the \ord[i] cyclic shift (i.e., $\sigma_i(j)=i+j$ if $i+j\le n$ and $i+j-n$ otherwise).
For $x\in\01^n$ and $\tau\in S_n$, denote by $\tau(x)$ the element of $\01^n$, whose \ord[\tau(i)] position contains $x_i$ for each $i$ -- in particular, $\sigma_j(x)$ is the \f j-bit cyclic shift of $x$.

For functions $f,g:\01^n\to\RR$, we define
\m{\ip fg
\deq 2^{-n}\tm\sum_{x\in\01^n}f(x)\tm g(x)
= \E[X\unin\01^n]{f(X)\tm g(X)}}
and $\norm[2] f \deq \sq{\ip ff}$.
For  $s\sbseq[n]$ and $x\in\01^n$, let $\chi_s(x)\deq(-1)^{\sz{x_s}}$ and $\hat f(s)\deq\ip f{\chi_s}$.
The \e{Fourier transform} $f\to\hat f$ is a \e{norm-preserving} linear mapping in the following sense:\
$\norm[2] f^2=\sum_s\hat f(s)^2$ (\e{Parseval's identity}).
The vectors $\chi_s$ form an orthonormal basis of $\RR[^{2^n}]$ and
\m{
f(x) = \sum_{s\sbseq[n]}\hat f(s)\tm\chi_s(x)
}
for every $x\in\01^n$.

\ndefi{small-bias spaces}{For $\eps\ge0$, we call $T\sbseq\01^n$ an \f\eps-bias space if
\m{
\sz{
\E[\tau\in T]{\chi_s(\tau)}
}\le\eps
}
for every $s\sbseq[n]$, $s\neq\emptyset$.
}

Being a small-bias space is a ``pseudorandom property'':\ it holds for random subsets of $\01^n$ almost always, and there are efficient constructions.

\nfct[f_s-bias]{\cite{NN93_Sm}}{For $\eps>0$, an \f\eps-bias space can be constructed deterministically in time $\poly(\dr n\eps)$.
Every pair of elements $\tau_1\neq\tau_2$ of the constructed space satisfies $\sz{\tau_1\+\tau_2}\in\fr n2\pm\aso n$.
}

The main communication problem studied in this work ($\ctEqT$) will be constructed using a small-bias space.
In order to argue the model separations, we do not need the definition of the problem to be explicit; nevertheless, we remark that our construction will be explicit in a rather strong sense:\ namely, $\ctEqT(x,y)$ will be computable in time $\poly(n)$ for any $x,y\in\01^n$.
This is due, in particular, to the complexity guarantees of \fctref{f_s-bias}.

\ssect{Communication complexity}

For an excellent survey of classical communication complexity, see~\cite{KN97_Comm}.
Quantum communication models differ from their classical counterparts in two aspects:\ the players are allowed to send quantum messages (accordingly, the complexity is measured in \e{qubits}) and to perform arbitrary quantum operations locally.

Of central importance to this work is the model of \e{simultaneous message passing (SMP)}, where there are 3 participants:\ \e{players} Alice and Bob, and \e{the referee}.
An SMP-protocol for computing a Boolean function $f(X,Y)$ has the following structure:
Alice receives $X$ and sends her message to the referee; at the same time, Bob receives $Y$ and sends his message to the referee; the referee uses the content of the two received messages to compute the answer.
The answer is correct when it equals $f(X,Y)$ (the input is always such that $f(X,Y)$ is defined).
We will consider the following variations of SMP:
\enum{
\item In \DII[\mu,\eps] (sometimes written as \DII[\eps] if $\mu$ is irrelevant or clear from the context) the players and the referee are \e{deterministic}, and the answer must be correct with probability at least $1-\eps$ when $(X,Y)\sim\mu$.\fn
{
In this work we will only deal with binary-valued functions; accordingly, we always assume that $\eps<\dr12$.
}
\item In \RII\ the players and the referee can use \e{local randomness}, and the answer must be correct with probability at least $\dr23$ for every valid input.
\item \RIIp\ is similar to \RII, but the players and the referee can use \e{shared randomness}.
\item In \QII\ the players can send \e{quantum} messages and the referee can apply any quantum measurement to compute the answer that must be correct with probability at least $\dr23$ for every valid input.
}

\sssect[sss_Eq]{Variations of equality}

The communication problem that we use for our separation is a function that can be viewed as a variation of the \e{equality} problem.

The \e{equality function} (viewed as a communication problem) is the following total\fn
{
A functional problem in communication complexity is called \e{total} when it is supported on the product set of the players' individual sets of input.
}
bipartite function.
Let $u\sbseq[n]$ (for technical reasons, we consider a ``projected version'' of equality), then
\m{
&\Equ:\01^{n+n}\to\01,\\
&\Equ(x,y)\deq\twocase{1}{if $x_u=y_u$;}{0}{otherwise.}
}
We write \Eq\ for $\Eq_{[n]}$.
Define input distributions for \Equ:
\itstart
\item for $a\in\01$, let $\mEqu[^a]$ be the uniform distribution over $\Equ[^{-1}](a)$;
\item let $\mEqu\deq\fr12\tm\l(\mEqu[^0]+\mEqu[^1]\r)$.
\itend

The next problem intuitively corresponds to asking whether $\Equ(X\+\tau,Y)=1$ for some $\tau$ from a predetermined set $T\sbseq\01^n$, usually of size $\poly(n)$ (in our analysis $T$ will be a small-bias space).
\m{
&\EquT:\01^{n+n}\to\01,\\
&\EquT(x,y)\deq\twocase{1}{if $(x\+\tau)_u=y_u$ for some $\tau\in T$;}{0}{otherwise.}
}
Define input distributions for \EquT:
\itstart
\item for $\tau\in T$, let $\mEqu[^\tau]$ be the distribution of $(X,Y)$ when $(X\+\tau,Y)\sim\mEqu$;
\item let $\mEquT\deq\fr1{|T|}\tm\sum_{\tau\in T}\mEqu[^\tau]$.
\itend

Next we define a ``noisy'' (or gapped) version of $\EqT$:
\m{
&\tEqT:\01^{n+n}\to\01;\\
&\tEqT(x,y)\deq
\fourcase
{1}{if $|x\+y\+\tau|\le\fr{6n}{15}$ for some $\tau\in T$}
{}{\tb and $|x\+y\+\tau|\nin(\fr{6n}{15},\fr{7n}{15})$ for every $\tau\in T$;}
{0}{if $|x\+y\+\tau|\ge\fr{7n}{15}$ for every $\tau\in T$;}
{\txt{undefined}}{otherwise.}
}

Intuitively, $\tEqT(x,y)$ ``asks'' whether $x\+\tau$ is close to $y$ with respect to one of the ``permitted'' bit-negations $\tau\in T$.
The \e{promise} is that $x\+\tau$ must be either far enough from $y$ (at distance $\ge\fr{7n}{15}$) or close to it (at distance $\le\fr{6n}{15}$) for every $\tau\in T$ -- otherwise the function is undefined.

Define input distributions for \tEqT:
\itstart
\item let $\mtEqT\deq\fr1{\chs n{\dr n3}}\tm\sum_{u\in\chs{[n]}{\dr n3}}\mEquT$.
\itend

We are ready to introduce the main communication problem considered in this work -- a function that can be viewed as a ``cyclic version'' of \EquT:
\m{
&\ctEqT:\01^{n+n}\to\01,\\
&\ctEqT(x,y)\deq
\fourcase
{1}
{if $|\sigma_j(x)\+y\+\tau|\le\fr{6n}{15}$ for some $\tau\in T$ and $j\in[n]$}
{}
{\tb and $|\sigma_j(x)\+y\+\tau|\nin(\fr{6n}{15},\fr{7n}{15})$ for every $\tau$ and $j$;}
{0}
{if $|\sigma_j(x)\+y\+\tau|\ge\fr{7n}{15}$ for every $\tau$ and $j$;}
{\txt{undefined}}
{otherwise.}
}

The intuition behind this definition is very similar to that behind $\tEqT(x,y)$, but the question here is whether $\sigma_j(x)+\tau\approx y$ with respect to some cyclic shift $\sigma_j$ and one of the bit-negations $\tau\in T$.

Define input distributions for \ctEqT:
\itstart
\item for $j\in[n]$, let $\mctEqT[^j]$ be the distribution of $(X,Y)$ when $(\sigma_j(X),Y)\sim\mtEqT$;
\item let $\mctEqT\deq\fr1n\tm\sum_{j\in[n]}\mctEqT[^j]$.
\itend

Let us also define the variants of our input distributions, where in the construction $\U[\01^{n+n}]$ replaces $\mEqu[^0]=\U[{\Equ[^{-1}](0)}]$.
For every $u\in\chs{[n]}{\dr n3}$, $\tau\in T$ and $j\in[n]$:
\itstart
\item let $\wbr\mEqu\deq\fr12\tm\l(\U[\01^{n+n}]+\mEqu[^1]\r)$.
\item let $\wbr{\mEqu[^\tau]}$ be the distribution of $(X,Y)$ when $(X\+\tau,Y)\sim\wbr\mEqu$;
\item let $\wbr\mEquT\deq\fr1{|T|}\tm\sum_{\tau\in T}\wbr{\mEqu[^\tau]}$.
\item let $\wbr\mtEqT\deq\fr1{\chs n{\dr n3}}\tm\sum_{u\in\chs{[n]}{\dr n3}}\wbr\mEquT$.
\item let $\wbr{\mctEqT[^j]}$ be the distribution of $(X,Y)$ when $(\sigma_j(X),Y)\sim\wbr\mtEqT$;
\item let $\wbr\mctEqT\deq\fr1n\tm\sum_{j\in[n]}\wbr{\mctEqT[^j]}$.
\itend
The above variants will be only used in the analysis, in which context they have a significant structural advantage:\ $\U[\01^{n+n}]$ is much more symmetric than $\mEqu[^0]$.
At the same time, these distributions are very close to their $\mEqu[^0]$-based originals, as formalised by the following claim.

\clm[cl_distr_l1]{
\m{
&\forall u\in\chs{[n]}{\dr n3},\,\tau\in T,\,j\in[n]:\\
&\tbb\norm[1]{\mEqu - \wbr\mEqu},\:
\norm[1]{\mEqu[^\tau] - \wbr{\mEqu[^\tau]}},\:
\norm[1]{\mEquT - \wbr\mEquT},\\
&\tbb\norm[1]{\mtEqT - \wbr\mtEqT},\:
\norm[1]{\mctEqT[^j] - \wbr{\mctEqT[^j]}},\:
\norm[1]{\mctEqT - \wbr\mctEqT} \, \in \, 2^{-\asOm n}
.}
}

The validity of the above claim follows readily by the Chernoff bound (\fctref{f_Cher}).

\sect[s_sum]{Intuition behind the new separation}

Recall that we are looking for a functional communication problem, easy for quantum but hard for classical SMP (naturally, equipped with shared randomness).
The initial inspiration comes from the observation that the most obvious quantum SMP protocol for \e{equality with gap (\tEq)} has certain ``robustness'' that seems impossible to achieve in a classical protocol.

Let
\m{
\tEq(x,y)\deq
\thrcase
{1}{if $|x\+y|\le\fr{n}{5}$;}
{0}{if $|x\+y|\ge\fr{2n}{5}$;}
{\txt{undefined}}{otherwise.}
}
A natural \QII-solution to this problem would be for Alice to send $\fr1{\sq n}\tm\sum_i\ket i\ket{X_i}$, for Bob to send $\fr1{\sq n}\tm\sum_i\ket i\ket{Y_i}$ and for the referee to perform the \e{swap test}~\cite{BCWW01_Qua} -- a quantum measurement with two possible outcomes, ``pass'' and ``fail'', where the probability of passing for states $\ket\alpha$ and $\ket\beta$ equals $\fr12+\fr{|\bket\alpha\beta|^2}2$.
In our case the passing probability is $\fr12+\fr{(n-|X\+Y|)^2}{2n^2}$, so estimating it with sufficient constant precision allows the referee to give the correct answer with constant-bounded error, thus solving the problem.\fn
{
For simplicity, in this informal overview we only require that a protocol solves a Boolean problem with error $\dr12-\asOm1$.
The definitions made in this part are not used elsewhere.
}

Note that the same pair of messages sent by the players can be used by the referee for solving
\m{
\tEq(\pi(X)\+\tau,Y)
} 
for \e{any} $\pi\in S_n$ and $\tau\in\01^n$:
Upon receiving the messages and before performing the swap test, the referee would have to apply the obvious unitary transformation to the message from Alice (namely, permuting the indices and negating some bit values).

Let $S\sbs S_n$, $T\sbs\01^n$ and $|S|,|T|\in\poly(n)$.
Using the above intuition, we conclude that there exists an efficient quantum protocol for the problem
\m{
\tEq[_{S,T}](x,y)\deq
\thrcase
{1}{if $|\pi(x)\+\tau\+y|\le\fr{n}{5}$ for some $\pi\in S$ and $\tau\in T$;}
{0}{if $|\pi(x)\+\tau\+y|\ge\fr{2n}{5}$ for every $\pi\in S$ and $\tau\in T$;}
{\txt{undefined}}{otherwise.}
}
To solve it in \QII, both Alice and Bob send \asO{\log n} copies of their messages from the \tEq-protocol described above, which allows the referee to solve any instance of $\tEq(\pi(X)\+\tau,Y)$ with error $1/poly(n)$ (arbitrarily small).
In particular, this means that he can ``reuse'' the messages and test $\tEq(\pi'(X)\+\tau',Y)$ for every $\pi'\in S$ and $\tau'\in T$ with polynomially-small error, thus solving the problem.

One can see that the main communication task studied here -- $\ctEqT$ -- is an instance of $\tEq[_{S,T}]$ with different constants, $S$ being the set of cyclic bit-shifts and $T$ being a small-bias space.

\ssect[ss_tow_low]{Towards the lower bound}

Proving a strong lower bound for the \RIIp-complexity of $\ctEqT$ is interesting for several reasons. 
One of them is the technical challenge:\ rather fine tuning of the method is required.

The bound has to \e{distinguish} between the models \RIIp\ and \RI, whose respective strengths are rather close to each other.
Indeed, not only $\ctEqT(X,Y)$, but any $\tEq[_{S,T}](X,Y)$ is easy for (randomised) one-way protocols:\ Alice can send to Bob a number of randomly selected pairs $(i,X_i)$, letting him estimate, with sufficient confidence and accuracy, the values of $|\pi(x)\+\tau\+y|$ for every $\pi\in S$ and $\tau\in T$.
Sending \asO{\log n} pairs for uniformly-chosen \pl[i] would suffice, and therefore $\RI(\ctEqT)\in\asO{\log^2(n)}$.

Ignoring some technical details, our lower-bound argument for $\RIIp(\ctEqT)$ can be outlined as follows.

First of all, we need a convenient characterisation of efficient protocols for \tEq.
It will be based on the observation that if a random input satisfying $X\approx Y$ is given to an \RIIp-protocol for $\tEq(X,Y)$, then the two messages received by the referee are likely to ``witness'' that fact.
After some technical manipulations, this idea will lead to
\m[m_intD]{
\E[i]{
{
\Delta_\alpha^{i}
\tm\Delta_\beta^{i}
}
}\in\asOm{\fr1n}
,}
where $\Delta_\alpha^{i}$ is the ``bias'' of the referee's knowledge about $X_i$, gained from Alice's message $Al(X)$, and $\Delta_\beta^{i}$ is defined similarly with respect to $Y$ and Bob's message $Bo(Y)$.

Next we take $T$ into account.
We will use its small-bias properties to conclude that a protocol for $\tEqT(X,Y)$ must satisfy
\m[m_intI]{
\E[i]
{
\I{X_i}{Al(X)}
\tm\I{Y_i}{Bo(Y)}
}\in\asOm{\fr1n}
.}

The bound in \bref{m_intI} is significantly stronger than that in \bref{m_intD}:
Both $X_i$ and $Y_i$ are uniformly-random bits, so ``bias'' $\gamma>0$ in the referee's knowledge, say, about $X_i$ corresponds to $\asT{\gamma^2}$ bits of information.
The ``quadratic improvement'' from \bref{m_intD} to \bref{m_intI} captures the ``added hardness'' in the transition from $\tEq$ to $\tEqT$ -- at least, from the point of view of our analysis.

Finally, we add cyclic shifts in order to ``disconnect'' $\I{X_i}{Al(X)}$ from $\I{Y_i}{Bo(Y)}$.
We will show that any protocol for $\ctEqT(X,Y)$ must satisfy
\m[m_intII]{
\E[i]{\I{X_i}{Al(X)}}
\tm \E[j]{\I{Y_j}{Bo(Y)}}
\in\asOm{\fr1n}
,}
and this gives the desired lower bound, as at least one of $\E[i]{\I{X_i}{Al(X)}}$ and $\E[j]{\I{Y_j}{Bo(Y)}}$ must be \asOm{\dr1{\sq n}} in order to satisfy \bref{m_intII}.

\sect[s_qua]{Solving \ctEqT\ with simultaneous quantum messages}

Here we construct a protocol for solving \ctEqT\ in \QII.
First we consider the following simpler problem.

For any $\tau\in T$ and $j\in[n]$, let
\m{
\tEq[_{j,\tau}](x,y)\deq
\thrcase
{1}{if $|\sigma_j(x)\+y\+\tau|\le\fr{6n}{15}$;}
{0}{if $|\sigma_j(x)\+y\+\tau|\ge\fr{7n}{15}$;}
{\txt{undefined}}{otherwise.}
}

\para{A protocol for \tEq[_{j,\tau}].}
Upon receiving the input, Alice and Bob send, respectively,
\m{
\ket{\phi_{Al}} \deq \fr1{\sq n}\tm\sum_{i=1}^n\ket i\ket{X_i}
\txt{~~and~~}
\ket{\phi_{Bo}} \deq \fr1{\sq n}\tm\sum_{i=1}^n\ket i\ket{Y_i}
}
to the referee.
The referee then applies $\sigma_{j}$ to the first register of $\ket{\phi_{Al}}$ and $\tau(\sigma_{j}(i))$-controlled negation to the second, thus transforming the state into
\m{
\ket{\phi_{Al}'}
\,=\,\fr1{\sq n}\tm\sum_{i=1}^n
\ket{\sigma_{j}(i)}\ket{X_i\+\tau\l(\sigma_{j}(i)\r)} 
.}
Note that the above transformation is orthogonal (in particular, reversible), and therefore can be performed, preserving the superposition.

At this point the referee can apply the swap test to the states $\ket{\phi_{Al}'}$ and $\ket{\phi_{Bo}}$, which would ``pass'' with probability
\m{
\fr{1+\sz{\bket{\phi_{Al}'}{\phi_{Bo}}}^2}2
=\fr12+\fr{(n-|\sigma_{j_0}(X)\+\tau_0\+Y|)^2}{2n^2}
\,\twocase
{>\fr23}{if $\tEq[_{j,\tau}](x,y)=1$;}
{<\fr{29}{45}}{if $\tEq[_{j,\tau}](x,y)=0$.}
}

For any $\eps>0$, let $\Cl P_{j,\tau}^\eps$ denote the protocol that repeats the above procedure \asO{\log\fr1\eps} times in parallel (in particular, the players send that many copies of, respectively, $\ket{\phi_{Al}}$ and $\ket{\phi_{Bo}}$), outputs ``$1$'' if at least $\fr{59}{90}$-fraction of the swap tests have passed and ``$0$'' otherwise -- the number of performed repetitions is chosen so that the resulting $\Cl P_{j,\tau}^\eps$ solves $\tEq[_{j,\tau}](X,Y)$ with error less than $\eps$.
The resulting communication cost of $\Cl P_{j,\tau}^\eps$ is \asO{\log n\tm\log\fr1\eps}.

Let $(\Pi_{j,\tau}^\eps, I-\Pi_{j,\tau}^\eps)$ be the \f2-outcome projective measurement that the referee applies in $\Cl P_{j,\tau}^\eps$ to the received messages in order to determine the answer (with the outcome $\Pi_{j,\tau}^\eps$ corresponding to answering ``$\tEq[_{j,\tau}](x,y)=1$''), and let this be the only step performed by the referee.\fn
{
Putting it differently, the measurement $(\Pi_{j,\tau}^\eps, I-\Pi_{j,\tau}^\eps)$ incorporates all the steps taken by the referee according to $\Cl P_{j,\tau}^\eps$.
}

Note that execution of $\Cl P_{j,\tau}^\eps$ doesn't require from either Alice or Bob the knowledge of either $j$ or $\tau$ -- only the referee has to know these values in order to apply $(\Pi_{j,\tau}^\eps, I-\Pi_{j,\tau}^\eps)$.
This makes $\Cl P_{i,\eps}$ a perfect ``building block'' for solving the original problem.

\para{A protocol for \ctEqT.}
Let Alice and Bob send their messages to the referee, as prescribed by $\Cl P_{j,\tau}^{\eps'}$ for some $\eps'$ to be fixed soon (recall that these messages do not depend on the values of $j$ and $\tau$).
The referee sequentially measures the received messages with $(\Pi_{j,\tau}^{\eps'}, I-\Pi_{j,\tau}^{\eps'})$ for all $\tau\in T$ and $j\in[n]$.
If at least one outcome $\Pi_{j,\tau}^{\eps'}$ has been obtained, the referee answers ``$\ctEqT(X,Y)=1$''; otherwise, ``$\ctEqT(X,Y)=0$''.

Call the above protocol $\Cl P$.
Assume without loss of generality that $\ctEqT(X,Y)\in\01$ (i.e., the input fulfils the promise).
To analyse the error of $\Cl P$, note that the protocol can return the wrong answer only if for some $(j,\tau)$ the outcome of the corresponding measurement $(\Pi_{j,\tau}^{\eps'}, I-\Pi_{j,\tau}^{\eps'})$ is \e{wrong} -- that is, the outcome is $\Pi_{j,\tau}^{\eps'}$ while $\tEq[_{j,\tau}](X,Y)=0$, or vice versa.
Note that while the probability of the outcome of the first performed measurement being wrong is bounded above by $\eps'$ (as follows trivially from the error bound of $\Cl P_{j,\tau}^{\eps'}$), at the subsequent rounds the state being measured may have been ``distorted'' by the earlier measurements, which, in turn, may increase the error probability.

It is known (e.g., see \fakelemref{2} in~\cite{A04_Lim}) that if a sequence of $m$ quantum measurements of the same state is performed, such that in every measurement the \e{most likely} outcome would occur with probability at least $1-\eps'$ if the measurement were performed on the ``clean'' state, then such $\eps'\in\poly(\dr1m)$ can be chosen, that all the $m$ obtained outcomes will be the most likely ones with probability at least $\dr23$ (or any other constant less than $1$).

For the protocol $\Cl P$ to be correct, it is enough for the measurement corresponding to every $\tau\in T$ and $j\in[n]$ to return the most likely value.
Accordingly, choosing $\eps'\in\dr1{\poly(n\tm|T|)}$ is sufficient for the resulting $\Cl P$ to solve $\ctEqT(X,Y)$ with error at most $\dr13$.
The respective protocol's communication complexity is, therefore, \asO{(\log n)^2+\log n\tm\log|T|}.

\crl[crl_ctEqT-upper]{For every $T\sbseq\01^n$,
\m{
\QII(\ctEqT)
\,\in\, \asO{(\log n)^2+\log n\tm\log|T|}
.}
}

\sect{A probabilistic interlude}

Here we prove several claims addressing the behaviour of non-independent random variables.
The statements are rather intuitive, though we are not aware of previously published proofs.

\ssect[ss_opt]{Optimistic inequalities}

\nclm[cl_optchain]{Optimistic chain inequality}
{Let $X_1\dc X_m$ be random variables, where each $X_i$ is supported on (finite) $G_i\cupdot B_i$.
Let $\mu$ denote the joint distribution of $X=(X_1\dc X_m)$, then
\m[m_optchain1]{
\PR[X\sim\mu]{\bigwedge_{j=1}^mX_j\in G_j}
&=\prod_{i=1}^m\PRr[X\sim\mu]{X_i\in G_i}{\bigwedge_{j=1}^{i-1}X_j\in G_j}\\
&=\prod_{i=1}^m
\Ee[X'\sim\mu]
{\PRr[X\sim\mu]{X_i\in G_i}{\bigwedge_{j=1}^{i-1}X_j=X_j'}}
{\bigwedge_{j=1}^{\textcolor{Red}{\bm{i-1}}}X_j'\in G_j}\\
&\le\prod_{i=1}^m
\Ee[X'\sim\mu]
{\PRr[X\sim\mu]{X_i\in G_i}{\bigwedge_{j=1}^{i-1}X_j=X_j'}}
{\bigwedge_{j=1}^{\textcolor{Red}{\bm m}}X_j'\in G_j}
,}
where $X'=(X_1'\dc X_m')$ and $X$ are independent from one another, unless conditioned explicitly.
Moreover,
\m[m_optchain2]{
&\Log{\PR[X\sim\mu]{\bigwedge_{j=1}^mX_j\in G_j}}\\
&\tbbb\le\sum_{i=1}^m
\Ee[X'\sim\mu]
{\Log{\PRr[X\sim\mu]{X_i\in G_i}{\bigwedge_{j=1}^{i-1}X_j=X_j'}}}
{\bigwedge_{j=1}^{m}X_j'\in G_j}
.}
}

The equalities in \bref{m_optchain1} correspond to the standard ``chain'' decomposition (included here for convenience).
In comparison to the standard decomposition, the inequality offers a more symmetric upper bound on $\PR{\wedge X_j\in G_j}$ at the expense of tightness.\fn
{
The statement of \clmref{cl_optchain} can probably be made tight via expressing the difference between the two sides of \bref{m_optchain2} as a sum of relative entropies, cf.\ \clmref{cl_optcond}.
}

We call the above claim \e{optimistic}, viewing the subsets $G_i$ as \e{good}, \pl[B_i] as \e{bad} and interpreting the statement of \bref{m_optchain1} as saying that \e{the estimated probability of $m$ good outcomes doesn't decrease as a result of making the estimation ``optimistically biased''}:\ instead of conditioning the expectation on $\lf[\bigwedge_{j=1}^{i-1}X_j'\in G_j\rt]$ (which would give the actual probability of all good outcomes), the right-hand side of the above inequality uses more ``good-oriented'' (and more restricting) condition $\lf[\bigwedge_{j=1}^{m}X_j'\in G_j\rt]$.

Moreover, the right-hand side of \bref{m_optchain2} \e{is likely to have grown as a result of making the expectations ``optimistically biased''} (i.e., conditioning it on $\lf[\bigwedge_{j=1}^{m}X_j'\in G_j\rt]$):\ due to the strict concavity of $\log$, the statement wouldn't hold if the condition $\lf[\bigwedge_{j=1}^{m}X_j'\in G_j\rt]$ were replaced by $\lf[\bigwedge_{j=1}^{i-1}X_j'\in G_j\rt]$, unless the quantities under the expectations are constant (that is, unless every event $\lf[X_i\in G_i\rt]$ is independent from the values of $X_1\dc X_{i-1}$, subject to $\lf[\bigwedge_{j=1}^{i-1}X_j\in G_j\rt]$).

Note also that the inequality in \bref{m_optchain1} isn't necessarily true ``element-wise'':\ there may exist a situation, where for some $i_0\in[m]$:
\m[P]{
\PRr[X\sim\mu]{X_{i_0}\in G_{i_0}}{\bigwedge_{j=1}^{i_0-1}X_j\in G_j}
&=\Ee[X'\sim\mu]
{\PRr[X\sim\mu]{X_{i_0}\in G_{i_0}}{\bigwedge_{j=1}^{i_0-1}X_j=X_j'}}
{\bigwedge_{j=1}^{i_0-1}X_j'\in G_j}\\
&>\Ee[X'\sim\mu]
{\PRr[X\sim\mu]{X_{i_0}\in G_{i_0}}{\bigwedge_{j=1}^{i_0-1}X_j=X_j'}}
{\bigwedge_{j=1}^mX_j'\in G_j}
.}
The following statement implies that the above inequality might hold only for $i_0<m$.

\nclm[cl_optcond]{Optimistic conditioning}
{Let $X_1$ and $X_2$ be random variables, each $X_i$ supported on (finite) $G_i\cupdot B_i$, and let $\mu$ be the joint distribution of $X=(X_1,X_2)$.
Then
\m{
&\log\l(\PRr[X\sim\mu]{X_2\in G_2}{X_1\in G_1}\r)
=\log\l(\Ee[X'\sim\mu]
{\PRr[X\sim\mu]{X_2\in G_2}{X_1=X_1'}}{X_1'\in G_1}\r)\\
&\tbbb
=\Ee[X'\sim\mu]
{\log\l(\PRr[X\sim\mu]{X_2\in G_2}{X_1=X_1'}\r)}{X_1'\in G_1,X_2'\in G_2}
-\KL\beta\alpha\\
&\tbbb
\le\Ee[X'\sim\mu]
{\log\l(\PRr[X\sim\mu]{X_2\in G_2}{X_1=X_1'}\r)}{X_1'\in G_1,X_2'\in G_2}
,}
where $X'=(X_1',X_2')$ is independent from $X$ (unless conditioned explicitly), and $\alpha$ and $\beta$ denote the distributions of $X_1$, conditioned, respectively, on $[X_1\in G_1]$ and on $[X_1\in G_1,X_2\in G_2]$.
}

The statement of \clmref{cl_optcond}, similarly to \bref{m_optchain2}, witnesses the qualitative ``benefit'' of optimistic conditioning:\ since $\log$ is strictly concave, whenever $[X_2\in G_2]$ depends on $X_1$ (subject to $[X_1\in G_1]$), the above inequality wouldn't hold if the expectation were not subject to $[X_2'\in G_2]$.

\prfstart[\clmref{cl_optcond}]
For every $c\in G_1$, let $p_c\deq\PR{X_1=c}$ and $q_c\deq\PRr{X_2\in G_2}{X_1=c}$.
Then
\m{
&\PR[X\sim\mu]{X_1\in G_1} = \sum_{d\in G_1} p_d,\\
&\alpha(c)=\PRr[X\sim\mu]{X_1=c}{X_1\in G_1} = \fr{p_c}{\sum_{d\in G_1} p_d},
}
and
\m{
&\PR[X\sim\mu]{X_1=c,X_2\in G_2} = p_cq_c,\\
&\PR[X\sim\mu]{X_1\in G_1,X_2\in G_2} = \sum_{d\in G_1} p_dq_d,\\
&\beta(c)=\PRr[X\sim\mu]{X_1=c}{X_1\in G_1,X_2\in G_2}
= \fr{p_cq_c}{\sum_{d\in G_1} p_dq_d}
.}

Let
\m{
k \deq \fr{\sum_{d\in G_1} p_dq_d}{\sum_{d\in G_1} p_d}
\= \fr{\alpha(c)}{\beta(c)}\tm q_c~~\txt{(for any $c\in G_1$)}
,}
then
\m{
\log\l(\PRr[X\sim\mu]{X_2\in G_2}{X_1\in G_1}\r)
&= \Log{\sum_{d\in G_1}\alpha(d)\tm q_d}
= \Log{\sum_{d\in G_1}k\tm\beta(d)}
= \log(k)\\
&= \sum_{d\in G_1}\beta(d)\tm\Log{k\tm\fr{\beta(d)}{\alpha(d)}}
- \sum_{d\in G_1}\beta(d)\tm\Log{\fr{\beta(d)}{\alpha(d)}}\\
&= \sum_{d\in G_1}\beta(d)\tm\Log{q_d}
- \KL\beta\alpha\\
&\hspace{-65pt}
= \Ee[X'\sim\mu]
{\log\l(\PRr[X\sim\mu]{X_2\in G_2}{X_1=X_1'}\r)}{X_1'\in G_1,X_2'\in G_2}
-\KL\beta\alpha
,}
as required (the stated inequality follows from the non-negativity of relative entropy).
\prfend[\clmref{cl_optcond}]

\prfstart[\clmref{cl_optchain}]
Let us first consider the case of two variables $(Y_1,Y_2)\sim\nu$, supported, respectively, on $\Cl G_1\cupdot\Cl B_1$ and $\Cl G_2\cupdot\Cl B_2$:
\m[m_Y1Y2]{
&\Log{\PR[\nu]{Y_1\in \Cl G_1,Y_2\in \Cl G_2}}\\
&\tbb=\Log{\PR{Y_1\in \Cl G_1}}
+ \Log{\PRr{Y_2\in \Cl G_2}{Y_1\in \Cl G_1}}\\
&\tbb\le\Log{\PR{Y_1\in \Cl G_1}}
+ \Ee[(Y_1',Y_2')\sim\nu]
{\Log{\PRr{Y_2\in \Cl G_2}{Y_1=Y_1'}}}
{Y_1'\in \Cl G_1,Y_2'\in \Cl G_2}
,}
as follows from \clmref{cl_optcond}.

Let $\mu'$ denote the distribution of $(X_1\dc X_m)\sim\mu$, conditioned upon $[\bigwedge_{j=1}^mX_j\in G_j]$; in other words,
\m{
\mu'(x_1\dc x_m)\deq\twocase
{\fr{\mu(x_1\dc x_m)}{\mu(G_1\dtimes G_m)}}{if $\bigwedge_{j=1}^mx_j\in G_j$;}
{0}{otherwise.}
}
Note that
\m[P]{
&\Log{\PR[X\sim\mu]{\bigwedge_{j=1}^mX_j\in G_j}}\\
&\tbb\le\Log{\PR[X\sim\mu]{X_1\in G_1}}
+\Ee[X'\sim\mu]
{\Log{\PRr[X\sim\mu]{\bigwedge_{j=2}^{m}X_j\in G_j}{X_1=X_1'}}}
{\bigwedge_{j=1}^mX_j'\in G_j}\\
&\tbb=\Log{\PR[X\sim\mu]{X_1\in G_1}}
+\E[X'\sim\mu']
{\Log{\PRr[X\sim\mu]{\bigwedge_{j=k+1}^{m}X_j\in G_j}{X_1=X_1'}}}
}
holds for $k=1$, as a direct application of \bref{m_Y1Y2}.

Inequality \bref{m_optchain2} follows by induction on $k$.~\fn
{
We could have started from the trivial case of $k=0$ and handle $k=1$ as a generic inductive step; we treat the latter as the base case in order to present the main idea behind the induction in a somewhat simpler form.
}
Assume that
\m[m_step]{
\Log{\PR[X\sim\mu]{\bigwedge_{j=1}^mX_j\in G_j}}
&\le \E[X'\sim\mu']{\sum_{i=1}^k\Log{\PRr[X\sim\mu]{X_i\in G_i}{\bigwedge_{j=1}^{i-1}X_j=X_j'}}}\\
&\tb+\E[X'\sim\mu']{
\Log{\PRr[X\sim\mu]{\bigwedge_{j=k+1}^{m}X_j\in G_j}{\bigwedge_{j=1}^{k}X_j=X_j'}}}\\
&=\E[X'\sim\mu']
{\sum_{i=1}^k\Log{\PRr[X\sim\mu]{X_i\in G_i}{\bigwedge_{j=1}^{i-1}X_j=X_j'}}}\\
&\tb+\sum_{x'\in\01^m}\mu'(x')\tm
\underbrace%
{\Log{\PRr[X\sim\mu]{\bigwedge_{j=k+1}^{m}X_j\in G_j}{\bigwedge_{j=1}^{k}X_j=x_j'}}}%
_{\textcolor{Red}\circledast}
}
holds for some $k\ge1$.

For any $x'\in\01^m$, let $\nu_{x'}^{(k)}$ denote the distribution of $\l(X_j\r)_{j=k+1}^{m}$ when $(X_j)_{j=1}^{m}\sim\mu$, conditioned upon $\bigwedge_{j=1}^{k}X_j=x_j'$; in other words,
\m{
\nu_{x'}^{(k)}(x_{k+1}\dc x_m)\deq
\fr{\mu(x_1'\dc x_k',x_{k+1}\dc x_m)}
{\sum_{x_{k+1}''\dc x_m''}\mu(x_1'\dc x_k',x_{k+1}''\dc x_m'')}
.}
Next we inspect $\textcolor{Red}\circledast$.
\m[m_*]{
\forall x':\:&\Log{\PRr[X\sim\mu]{\bigwedge_{j=k+1}^{m}X_j\in G_j}{\bigwedge_{j=1}^{k}X_j=x_j'}}
=\Log{\PR[X\sim\nu_{x'}^{(k)}]{\bigwedge_{j=k+1}^{m}X_j\in G_j}}\\
&\le\Log{\PR[X\sim\nu_{x'}^{(k)}]{X_{k+1}\in G_{k+1}}}\\
&\tb+\Ee[X''\sim\nu_{x'}^{(k)}]
{\Log{\PRr[X\sim\nu_{x'}^{(k)}]
{\bigwedge_{j=k+2}^{m}X_j\in G_j}{X_{k+1}=X_{k+1}''}}}
{\bigwedge_{j=k+1}^mX_j''\in G_j}
,}
where ``$X\sim\nu_{x'}^{(k)}$'' stands for $(X_{k+1}\dc X_m)\sim\nu_{x'}^{(k)}$, $X''=(X_{k+1}''\dc X_m'')$ and the inequality is an application of \bref{m_Y1Y2}.

Consider the following distribution.
\itemi{
\item Let $X'\sim\mu'$; denote by $x'$ the value taken by $X'$;
\item let $X''\sim\nu_{x'}^{(k)}$, subject to $\lf[\bigwedge_{j=k+1}^mX_j''\in G_j\rt]$.
}
We claim that the resulting distribution of $(X_1'\dc X_k',X_{k+1}''\dc X_m'')$ is simply $\mu'$:
\m[P]{
&\forall (x_1\dc x_m)\in G_1\dtimes G_m:\\
&\tb\mu'(x_1\dc x_m) = \fr{\mu(x_1\dc x_m)}{\mu(G_1\dtimes G_m)};\\
&\tb\PR{(X_1'\dc X_k')=(x_1\dc x_k)}=
\sum_{(x_{k+1}''\dc x_m'')\in G_{k+1}\dtimes G_m}
\fr{\mu(x_1\dc x_k,x_{k+1}''\dc x_m'')}{\mu(G_1\dtimes G_m)}.\\
&\forall (x_{k+1}'\dc x_m')\in G_{k+1}\dtimes G_m:\\
&\tb\PR{(X_1'\dc X_k',X_{k+1}''\dc X_m'')=(x_1\dc x_k,x_{k+1}'\dc x_m')}\\
&\tbb=\PR{(X_1'\dc X_k')=(x_1\dc x_k)} \tm
\fr{\nu_{x_1\dc x_k}^{(k)}(x_{k+1}'\dc x_m')}
{\nu_{x_1\dc x_k}^{(k)}(G_{k+1}\dtimes G_m)}\\
&\tbb=\fr
{\PR{(X_1'\dc X_k')=(x_1\dc x_k)} \tm \nu_{x_1\dc x_k}^{(k)}(x_{k+1}'\dc x_m')}
{\PRr[X\sim\mu]{\bigwedge_{j=k+1}^mX_j\in G_j}{\bigwedge_{j=1}^kX_j=x_j}};\\
&\tb\PR{(X_1'\dc X_k')=(x_1\dc x_k)}
\tm \nu_{x_1\dc x_k}^{(k)}(x_{k+1}'\dc x_m')\\
&\tbb=
\fr{\sum_{(x_{k+1}''\dc x_m'')\in G_{k+1}\dtimes G_m}
\mu(x_1\dc x_k,x_{k+1}''\dc x_m'')}
{\sum_{x_{k+1}''\dc x_m''}\mu(x_1\dc x_k,x_{k+1}''\dc x_m'')}
\tm \fr{\mu(x_1\dc x_k,x_{k+1}'\dc x_m')}{\mu(G_1\dtimes G_m)}\\
&\tbb=
\fr{\PR[X\sim\mu]{\bigwedge_{j=1}^kX_j=x_j \wedge \bigwedge_{j=k+1}^mX_j\in G_j}}
{\PR[X\sim\mu]{\bigwedge_{j=1}^kX_j=x_j}}
\tm \mu'(x_1\dc x_k,x_{k+1}'\dc x_m')\\  
&\tbb=\PRr[X\sim\mu]{\bigwedge_{j=k+1}^mX_j\in G_j}{\bigwedge_{j=1}^kX_j=x_j}
\tm \mu'(x_1\dc x_k,x_{k+1}'\dc x_m');\\
&\tb\PR{(X_1'\dc X_k',X_{k+1}''\dc X_m'')=(x_1\dc x_k,x_{k+1}'\dc x_m')}\\  
&\tbb=\mu'(x_1\dc x_k,x_{k+1}'\dc x_m')
,}
where we have somewhat abused the notation by writing ``$\nu_{x_1\dc x_k}^{(k)}$'' (note that $\nu_{x'}^{(k)}$ indeed depends only on the first $k$ bits of $x'$).

Accordingly, it follows from \bref{m_*} and from the definition of $\nu_{x'}^{(k)}$ that
\m[P]{
&\sum_{x'\in\01^m}\mu'(x')\tm\underbrace%
{\Log{\PRr[X\sim\mu]{\bigwedge_{j=k+1}^{m}X_j\in G_j}{\bigwedge_{j=1}^{k}X_j=x_j'}}}%
_{\textcolor{Red}\circledast}\\
&\tb\le\sum_{x'}\mu'(x')\tm\Log{\PR[X\sim\nu_{x'}^{(k)}]{X_{k+1}\in G_{k+1}}}\\
&\tbb+\sum_{x'}\mu'(x')\tm\Ee[X''\sim\nu_{x'}^{(k)}]
{\Log{\PRr[X\sim\nu_{x'}^{(k)}]{\bigwedge_{j=k+2}^{m}X_j\in G_j}{X_{k+1}=X_{k+1}''}}}
{\bigwedge_{j=k+1}^mX_j''\in G_j}\\
&\tb=\sum_{x'}\mu'(x')\tm
\Log{\PRr[X\sim\mu]{X_{k+1}\in G_{k+1}}{\bigwedge_{j=1}^{k}X_j=x_j'}}\\
&\tbb+\sum_{x'}\mu'(x')\\
&\tbb~\tm\Ee[X''\sim\nu_{x'}^{(k)}]
{\Log{\PRr[X\sim\mu]{\bigwedge_{j=k+2}^{m}X_j\in G_j}
{\bigwedge_{j=1}^{k}X_j=x_j' \wedge X_{k+1}=X_{k+1}''}}}
{\bigwedge_{j=k+1}^mX_j''\in G_j}\\
&\tb=\E[X'\sim\mu']
{\Log{\PRr[X\sim\mu]{X_{k+1}\in G_{k+1}}{\bigwedge_{j=1}^{k}X_j=X_j'}}}\\
&\tbb+\E[X''\sim\mu']
{\Log{\PRr[X\sim\mu]{\bigwedge_{j=k+2}^{m}X_j\in G_j}
{\bigwedge_{j=1}^{k+1}X_j=X_j''}}}
.}
Substituting it to \bref{m_step} gives
\m[P]{
\Log{\PR[X\sim\mu]{\bigwedge_{j=1}^mX_j\in G_j}}
&\le\E[X'\sim\mu']
{\sum_{i=1}^k\Log{\PRr[X\sim\mu]{X_i\in G_i}{\bigwedge_{j=1}^{i-1}X_j=X_j'}}}\\
&\tbb+\E[X'\sim\mu']
{\Log{\PRr[X\sim\mu]{X_{k+1}\in G_{k+1}}{\bigwedge_{j=1}^{k}X_j=X_j'}}}\\
&\tbb+\E[X'\sim\mu']
{\Log{\PRr[X\sim\mu]{\bigwedge_{j=k+2}^{m}X_j\in G_j}
{\bigwedge_{j=1}^{k+1}X_j=X_j'}}}\\
&=\E[X'\sim\mu']
{\sum_{i=1}^{k+1}\Log{\PRr[X\sim\mu]{X_i\in G_i}{\bigwedge_{j=1}^{i-1}X_j=X_j'}}}\\
&\tbb+\E[X'\sim\mu']
{\Log{\PRr[X\sim\mu]{\bigwedge_{j=k+2}^{m}X_j\in G_j}
{\bigwedge_{j=1}^{k+1}X_j=X_j'}}}
,}
thus completing the induction step; for $k=m-1$ the above reads:
\m{
\Log{\PR[X\sim\mu]{\bigwedge_{j=1}^mX_j\in G_j}}
& \le\E[X'\sim\mu']
{\sum_{i=1}^m\Log{\PRr[X\sim\mu]{X_i\in G_i}{\bigwedge_{j=1}^{i-1}X_j=X_j'}}}\\
& = \Ee[X'\sim\mu]
{\sum_{i=1}^m\Log{\PRr[X\sim\mu]{X_i\in G_i}{\bigwedge_{j=1}^{i-1}X_j=X_j'}}}
{\bigwedge_{j=1}^{m}X_j'\in G_j}
,}
which is precisely \bref{m_optchain2}; \bref{m_optchain1} follows by the concavity of $\log$.
\prfend[\clmref{cl_optchain}]

As a side note, we give the following generalisation, where the \ord[i] ``goodness criterion'' may depend not only on the value taken by $X_i$, but also on the values of $X_1\dc X_{i-1}$, as long as the condition is ``monotone non-increasing'' (e.g., the value of $(X_1,X_2)$ cannot be good when that of $X_1$ is bad).

\crl[crl_optchain]
{Let $X_1\dc X_m$ be random variables, so that for each $i\in[n]$ the tuple $(X_j)_{j=1}^i$ is supported on (finite) $\Cl G_i\cupdot\Cl B_i$ and for all $i_1<i_2$ it holds that $\Cl G_{i_2}$ projected to the first $i_1$ coordinates is a subset of $\Cl G_{i_1}$.
Let $\mu$ denote the joint distribution of $X=(X_1\dc X_m)$, then
\m{
\PR[X\sim\mu]{X\in\Cl G_m}
\le\prod_{i=1}^m
\Ee[X'\sim\mu]
{\PRr[X\sim\mu]{(X_j)_{j=1}^i\in\Cl G_i}{\bigwedge_{j=1}^{i-1}X_j=X_j'}}
{X'\in\Cl G_m}
,}
where $X'=(X_1'\dc X_m')$ and $X$ are independent from one another, unless conditioned explicitly.
}

\prfstart
For every $i\in[m]$, let $Y_i$ be a random variable that takes value $(X_i,Q_i)$, where
\m{
Q_i=\twocase{1}{if $(X_j)_{j=1}^i\in\Cl G_i$;}{0}{otherwise.}
}
Let $G_i\deq\supp(X_i)\times\set1$ and apply \clmref{cl_optchain} to the case of random variables $Y_i$ and ``good'' sets $G_i$.
\prfend

\ssect[ss_pred]{Confidence-weighted accuracy of Boolean prediction}

\clm[cl_pred]{Let $X$ and $Y$ be random variables, $X$ being supported on $\01$.
Then
\m{
\E[X',Y']{\PRr[X,Y]{X=X'}{Y=Y'}} - \fr12
\,=\, 2\tm\E[Y=y]{\l(\Ee[X]X{Y=y}-\fr12\r)^2}
,}
where $(X',Y')$ are distributed independently and identically to $(X,Y)$.

In particular, if $X\sim\U[\01]$, then
\m{
\E[X',Y']{\PRr[X,Y]{X=X'}{Y=Y'} - \fr12}
\in\asT{\I XY}
.}
}

Intuitively, if we view $X$ as \e{unknown}, $Y$ as \e{known}, and try to predict the former using the latter, then the expectation of $\PRr{X=X'}{Y=Y'}-\dr12$ can be interpreted as \e{confidence-weighted accuracy} when $X\sim\U[\01]$.\fn
{
Interpret the pair $(X',Y')$ as the ``actual outcome'' of the experiment, then $\PRr{X=X'}{Y=Y'}$ measures ``how likely'' the value of $X$ was to equal $X'$, conditioned upon the value of $Y$ being $Y'$.
}
It can be opposed to the standard notion of \e{confidence}:
\m{
\E[Y=y]{\sz{\Ee[X]X{Y=y} - \fr12}}
=\E[X',Y']{\color{Red}\lf|\color{black}\PRr[X,Y]{X=X'}{Y=Y'} - \fr12\color{Red}\rt|}
\in\asT{\sq{\I XY}}
.}
The qualitative difference between the two quantities is witnessed by the claim.

\prfstart[\clmref{cl_pred}]
Let $g(y)\deq\PRr{X=0}{Y=y}$ for every $y\in\supp(Y)$, then
\m{
&\hspace{-35pt}
\E[X',Y']{\PRr[X,Y]{X=X'}{Y=Y'}}\\
&=\E[Y'=y']
{\PRr{X'=0}{Y'=y'}\tm g(y') + \PRr{X'=1}{Y'=y'}\tm (1-g(y'))}\\
&=\E[Y']{g(Y')\tm g(Y')+(1-g(Y'))\tm(1-g(Y'))}\\
&=2\tm\E[Y]{\l(g(Y)-\fr12\r)^2}+\fr12
\,=\, 2\tm\E[Y=y]{\l(\Ee[X]X{Y=y}-\fr12\r)^2}+\fr12
.}

If $X\sim\U[\01]$, then
\m{
\h X-\hh X{Y=y}
= 1-\hh X{Y=y}
\in\asT{\l(\Ee X{Y=y}-\fr12\r)^2}
}
for every $y\in\supp(Y)$, and therefore,
\m{
\I XY =\E[Y=y]{\h X-\hh X{Y=y}}
\in\asT{\E[Y=y]{\l(\Ee[X]X{Y=y}-\fr12\r)^2}}
,}
as required.
\prfend

\sect[s_cla]{The \RIIp-complexity of \ctEqT\ -- a lower bound}

\ndefi[d_pDII]{protocols in \DII[\eps]}{
Let $\Cl P$ be a protocol in \DII[\eps], where both Alice and Bob send $r$ bits to the referee.
\itstart
\item Let $Al:\01^n\to\01^r$ be the ``message function'' of Alice, according to $\Cl P$ -- i.e., $Al(x)$ is sent when she receives input $x$;
\item let $\alpha:\01^n\to\Pow(\01^n)$ be the ``neighbourhood function'' corresponding to $Al(\dt)$ -- i.e., $\alpha(x)\deq\sett{x'}{Al(x')=Al(x)}$;
\item define $Bo(y)$ and $\beta(y)$ similarly.
\itend
}

Note that $\alpha(\dt)$ and $\beta(\dt)$ naturally correspond to \e{partitions} of, respectively, Alice's and Bob's input spaces:\ every possible message sent by a player corresponds to an element of his partition which is the set of input values corresponding to this message.
These partitions are fully determined by the message functions $Al(\dt)$ and $Bo(\dt)$ and, in some sense, they reveal ``all that matters'' about a protocol in \DII[\mu,\eps], as we can always consider (in the context of lower bounds, \e{assume}) an ``optimal'' referee -- the one who outputs a most likely guess regarding $f(X,Y)$ with respect to $\mu$, given the messages $Al(X)$ and $Bo(Y)$ from the players.

To analyse the complexity of \ctEqT, we reason as follows.
\itstart
\item We identify a useful property of all sufficiently accurate protocols for \Equ\ (cf.~\crlref{crl_Eq}).
\item We consider protocols for \EquT\ and see that a more rigid form of the above property must hold if $T$ is a so-called ``small-bias space'' (cf.~\lemref{l_EqT}).
\item We view \tEqT\ as ``\EquT\ on a random subset $u$'' -- accordingly, a protocol for \tEqT\ must satisfy the above characterisation with respect to ``random projections'', which leads to a more symmetric criterion (cf.~\lemref{l_tEqT}).
\item We observe that a protocol for \ctEqT\ must, in a sense, simultaneously solve $n$ ``rotated instances'' of \tEqT\ -- therefore, such a protocol must satisfy the $n$ ``rotated versions'' of the above characterisation, which in turn leads to an even more symmetric criterion (cf.~\lemref{l_ctEqT}) and then to the desired complexity lower bound (cf.~\crlref{crl_ctEqT-low}).
\itend

\ssect{Characterising protocols for \Equ}

To characterise protocols that solve the equality problem, we use the following idea:
Suppose for simplicity that $u=[n]$ (i.e., the protocol solves the standard \Eq).
If the partitions of $\01^n$ defined by $\alpha(\dt)$ and $\beta(\dt)$ are suitable for solving \Eq, then with respect to $X=Y\unin\01^n$, the pair of subsets $(\alpha(X),\beta(Y))$ will (typically) be such that $[X=Y]$ is ``likely'', given the messages -- namely,
\m{\PR[(X',Y')\unin\alpha(X)\times\beta(X)]{X'=Y'}
\gg\PR[(X',Y')\unin\01^{n+n}]{X'=Y'}=\fr1{2^n}.}
Applying the optimistic chain inequality (\clmref{cl_optchain}) with respect to the event $[X'=Y']=[\bigwedge_iX_i'=Y_i']$ and integrating over the rectangles of the form $\alpha(x)\times\beta(x)$ will lead to a convenient protocol characterisation.

\ndefi[d_pEqu]{protocols for \Equ}{
Fix some $u\sbseq[n]$ and let $\Cl P$ be a protocol that solves \Equ\ in \DII[\mEqu,\eps].
In addition to $Al(\dt)$, $Bo(\dt)$, $\alpha(\dt)$ and $\beta(\dt)$ defined earlier, we will use the following variations:
Let $z\in\01^{|u|}$, then
\itstart
\item denote by $Al^*(z)$ the \e{distribution} over $\01^r$, corresponding to $Al(X')$ when $X'$ is chosen uniformly at random from $\sett{x'\in\01^n}{x_u'=z}$;
\item denote by $\alpha^*(z)$ the \e{distribution} over $\Pow(\01^n)$, corresponding to $\sett{x'}{Al(x')=m_0}$ when $m_0$ is the value taken by $M\sim Al^*(z)$ (alternatively, $\alpha^*(z)$ can be defined as the distribution of $\alpha(X')$ when $X'$ is chosen uniformly at random from $\sett{x'\in\01^n}{x_u'=z}$);
\item define $Bo^*(z)$ and $\beta^*(z)$ similarly.
\itend
}

We will argue that the following type of objects are, in a sense, ``typical for $\Cl P$'' (that will be the technical core of our characterisation).

\ndefi[d_good]{good rectangles}{Let $A, B\sbseq\01^n$.
We call the rectangle $A\times B\sbseq\01^{n+n}$ good if
\m{
\PR[(X',Y')\unin A\times B]{X_u'=Y_u'}
\ge \fr1{4\sq\eps+2^{-|u|}}\tm\fr1{2^{|u|}}
.}
}

Our first step in this part is characterising good rectangles in a technically-convenient manner.
We need the following.

\ndefi[d_delta]{delta-properties of sets and partitions}{
Let $W\sbseq\01^n$, $i\in[|u|]$ and $z\in\01^{|u|}$.
Then
\m{
&\delta_W^{u,i}(z)
\deq\PRr[X\unin W]{X_u(i)=z_i}{X_u([i-1])=z_{[i-1]}}-\fr12,\\
&\Delta_\alpha^{u,i}(z)
\deq \PRr[X\unin \alpha^*(z)]{X_u(i)=z_i}{X_u([i-1])=z_{[i-1]}}-\fr12
\tb\lf\{=\E[A\sim\alpha^*(z)]{\delta_A^{u,i}(z)}\rt\},
}
and similarly for $\Delta_\beta^{u,i}(z)$.
}

\lem[l_good-cha]{Let $A, B\sbseq\01^n$.
If the rectangle $A\times B$ is good, then
\m{
\E[Z]{\sum_{i=1}^{|u|}\delta_A^{u,i}(Z)\tm\delta_B^{u,i}(Z)}
\ge \fr14\tm \ln\l( \fr1{4\sq\eps+2^{-|u|}} \r)
,}
where $Z$ is distributed as $X_u$ when $(X,Y)\unin A\times B$ conditioned on $[X_u=Y_u]$.
}

\prfstart
By the definition of good rectangles,
\m{
\fr1{4\sq\eps+2^{-|u|}}\tm\fr1{2^{|u|}}
&\le\PR[(X',Y')\unin A\times B]{X_u'=Y_u'}
= \PR{\bigwedge_{i=1}^{|u|}X_u'(i)=Y_u'(i)}\\
&\le \prod_{i=1}^{|u|} \Ee[(X',Y')\unin A\times B]
{~\textcolor{Red}\circledast~}{X_u'=Y_u'}\\
&= \prod_{i=1}^{|u|} \E[Z]
{
\PRr[(X,Y)\unin A\times B]{X_u(i)=Y_u(i)}
{X_u([i-1])=Y_u([i-1])=Z_{[i-1]}}
}
,}
where the second inequality is the optimistic chain inequality (\clmref{cl_optchain}), $\textcolor{Red}\circledast$ stands for
\m{
\PRr[(X,Y)\unin A\times B]{X_u(i)=Y_u(i)}
{X_u([i-1])=X_u'([i-1]),\,Y_u([i-1])=Y_u'([i-1])}
}
and $Z$ is distributed as $X_u'$ when $(X',Y')\unin A\times B$ conditioned on $[X_u'=Y_u']$.

On the other hand, for every $i\in[|u|]$ and $z\in\01^{|u|}$:
\m{
&\PRr[(X,Y)\unin A\times B]{X_u(i)=Y_u(i)}{X_u([i-1])=Y_u([i-1])=z_{[i-1]}}\\
&=\PRr{X_u(i)=Y_u(i)=z_i}{X_u([i-1])=Y_u([i-1])=z_{[i-1]}}\\
&\hspace{7pt}
+\PRr{X_u(i)=Y_u(i)=1-z_i}{X_u([i-1])=Y_u([i-1])=z_{[i-1]}}\\
&=\PRr[X\unin A]{X_u(i)=z_i}{X_u([i-1])=z_{[i-1]}}
\tm\PRr[Y\unin B]{Y_u(i)=z_i}{Y_u([i-1])=z_{[i-1]}}\\
&\hspace{7pt}
+\PRr[X\unin A]{X_u(i)=1-z_i}{X_u([i-1])=z_{[i-1]}}
\tm\PRr[Y\unin B]{Y_u(i)=1-z_i}{Y_u([i-1])=z_{[i-1]}}\\
&=\fr12 + 2\tm
\l(\PRr[X\unin A]{X_u(i)=z_i}{X_u([i-1])=z_{[i-1]}}-\fr12\r)\\
&\hspace{85pt}
\tm\l(\PRr[Y\unin B]{Y_u(i)=z_i}{Y_u([i-1])=z_{[i-1]}}-\fr12\r)\\
&=\fr12 + 2 \tm \delta_A^{u,i}(z) \tm \delta_B^{u,i}(z)
.}
Therefore,
\m{
\fr1{4\sq\eps+2^{-|u|}}\tm\fr1{2^{|u|}}
\le \prod_{i=1}^{|u|}
\l(
\fr12 + 2 \tm
\E[Z]{\delta_A^{u,i}(Z) \tm \delta_B^{u,i}(Z)}
\r)
,}
where $Z$ is distributed as $X_u$ when $(X,Y)\unin A\times B$ conditioned on $[X_u=Y_u]$.

So,
\m{
\ln\l(\fr1{4\sq\eps+2^{-|u|}}\tm\fr1{2^{|u|}}\r)
&\le \sum_{i=1}^{|u|}
\l(
\ln\l(\fr12\r)
+\ln\l(1+ 4\tm \E[Z]{\delta_A^{u,i}(Z) \tm \delta_B^{u,i}(Z)}\r)
\r)\\
&\le |u|\tm\ln\l(\fr12\r)
+4\tm\sum_{i=1}^{|u|} \E[Z]{\delta_A^{u,i}(Z) \tm \delta_B^{u,i}(Z)}
,}
as required.
\prfend[\lemref{l_good-cha}]

Next we will ``look inside'' $\Cl P$, for which we need the following.

\ndefi[d_ravaXY]{random variables corresponding to $[X_u=Y_u]$}{
\itstart
\item Let $Z\sim\U[\01^{|u|}]$.
\item Let the pair of $\Pow(\01^n)$-valued variables $(\Cl A,\Cl B)$ be distributed as $(\alpha^*(Z),\beta^*(Z))$.
\item Let $Z'$ be distributed as $X_u$ when $(X,Y)\unin \Cl A\times \Cl B$ conditioned on $[X_u=Y_u]$.
\itend
}

Intuitively, the variable $Z$ corresponds to sampling the protocol input from $\mEqu[^1]$:\ think of it as drawing uniformly-random $(X,Y)$, subject to $X_u=Y_u=Z$.
Then the rectangle $\Cl A\times\Cl B$ can be viewed as the knowledge that the referee obtains from the players' messages regarding the input pair.
View $Z'$ as a ``sibling of $Z$'', used in the proof for technical reasons.

Note two Markov chains that correspond to these random variables:
\m{
\Cl A\leftrightarrow Z\leftrightarrow\Cl B
\tb\txt{and}\tb
Z\leftrightarrow (\Cl A,\Cl B)\leftrightarrow Z'
,}
in other words, $A$ and $B$ are independent when conditioned on $Z$, and $Z$ and $Z'$ are independent when conditioned on $(\Cl A,\Cl B)$.

We claim that the latter chain is \e{symmetric} in the following sense:

\lem[l_sym]{The marginal distributions of $((\Cl A,\Cl B), Z)$ and of $((\Cl A,\Cl B), Z')$ are the same.}

\prfstart
Let $(\Cll a,\Cll b)\in\supp(\Cl A,\Cl B)$ and denote by $[(\Cll a,\Cll b)]$ the event that $(\Cl A,\Cl B)=(\Cll a,\Cll b)$, by $[\Cll a]$ the event that $\Cl A=\Cll a$ and by $[\Cll b]$ the event that $\Cl B=\Cll b$.
Let $z_0\in\01^{|u|}$, then
\m{
\PRr{(\Cll a,\Cll b)}{Z=z_0}
&=\PRr{\Cll a}{Z=z_0}\tm\PRr{\Cll b}{Z=z_0}\\
&=\PRr{Z=z_0}{\Cll a}
\tm\fr{\PR{\Cll a}}{\PR{Z=z_0}}
\tm\PRr{Z=z_0}{\Cll b}
\tm\fr{\PR{\Cll b}}{\PR{Z=z_0}}\\
&=\PRr{Z=z_0}{\Cll a}
\tm\PR{\Cll a}
\tm\PRr{Z=z_0}{\Cll b}
\tm\PR{\Cll b}
\tm2^{2|u|}
.}
On the other hand,
\m{
&\PR{\Cll a}
= \PR[Z\unin\01^{|u|}]{\alpha^*(Z)=\Cll a}
= \PR[X\unin\01^n]{\alpha(X)=\Cll a}
= \fr{|\Cll a|}{2^n},\\
&\PRr{Z=z_0}{\Cll a}
= \PRr{Z=z_0}{\alpha^*(Z)=\Cll a}
= \PRr{X_u=z_0}{\alpha(X)=\Cll a}
= \PR[X\unin\Cll a]{X_u=z_0}
,}
and similarly for $\PR{\Cll b}$ and $\PRr{Z=z_0}{\Cll b}$.
Accordingly,
\m{
\PRr{(\Cll a,\Cll b)}{Z=z_0}
= \PR[X\unin\Cll a]{X_u=z_0} \tm \PR[Y\unin\Cll b]{Y_u=z_0}
\tm|\Cll a| \tm|\Cll b| \tm2^{2|u|-2n}
.}
Therefore,
\m[m_abz]{
\PR{[(\Cll a,\Cll b)]\wedge Z=z_0}
&= \PR{Z=z_0}\tm\PRr{(\Cll a,\Cll b)}{Z=z_0}\\
&= \PR[X\unin\Cll a]{X_u=z_0} \tm \PR[Y\unin\Cll b]{Y_u=z_0}
\tm|\Cll a| \tm|\Cll b| \tm2^{|u|-2n}
}
and
\m[m_ab]{
\PR{(\Cll a,\Cll b)}
&=\sum_z\PR{Z=z}
\tm\PR[X\unin\Cll a]{X_u=z} \tm \PR[Y\unin\Cll b]{Y_u=z}
\tm|\Cll a| \tm|\Cll b| \tm2^{2|u|-2n}\\
&=\PR[\mac{X\unin\Cll a\\Y\unin\Cll b}]{X_u=Y_u}
\tm|\Cll a| \tm|\Cll b| \tm2^{|u|-2n}
.}

On the other hand,
\m{
\PR{[(\Cll a,\Cll b)]\wedge Z'=z_0}
&=\PRr{Z'=z_0}{(\Cll a,\Cll b)} \tm \PR{(\Cll a,\Cll b)}\\
&=\fr{\PR[X\unin\Cll a]{X_u=z_0} \tm \PR[Y\unin\Cll b]{Y_u=z_0}}
{\PR[\mac{X\unin\Cll a\\Y\unin\Cll b}]{X_u=Y_u}}
\tm \PR{(\Cll a,\Cll b)}\\
&=\PR[X\unin\Cll a]{X_u=z_0} \tm \PR[Y\unin\Cll b]{Y_u=z_0}
\tm|\Cll a| \tm|\Cll b| \tm2^{|u|-2n}\\
&=\PR{[(\Cll a,\Cll b)]\wedge Z=z_0}
,}
where the last two inequalities follow from \bref{m_ab} and \bref{m_abz}, respectively.
\prfend[\lemref{l_sym}]

Our characterisation of $\Cl P$ will be based on the following structural observation.

\lem[l_Eq-good]{
\m{
\PR[\Cl A,\Cl B]{\txt{$\Cl A\times\Cl B$ is a good rectangle}}
> 1-2\eps-2\sq\eps
.}
}

\prfstart
Let $(a,b)\in\01^{r+r}$ be a pair of players' messages and
\m{
err(a,b)\deq
\PRr[(X,Y)\sim\mEqu]{\txt{$\Cl P(X,Y)$ makes an error}}{Al(X)=a,Bo(Y)=b}
.}
By the correctness assumption,
\m{
\PR[(X,Y)\sim\mEqu]{err(Al(X),Bo(Y))>\sq\eps}<\sq\eps
.}
Call a pair of messages $(a,b)\in\01^{r+r}$ \e{bad} if $err(a,b)>\sq\eps$ and \e{good} otherwise.

Recall that $\mEqu$ is the ``uniform mixture'' of $\mEqu[^0]$ and $\mEqu[^1]$.
Accordingly, from the correctness assumption it follows that with respect to $(X,Y)\sim\mEqu[^1]$,
\itstart
\item $\Cl P$ accepts (that is, produces output ``$1$'') with probability at least $1-2\eps$;
\item $(Al(X),Bo(Y))$ is a bad message with probability at most $2\sq\eps$.
\itend
Note that sampling $(Al(X),Bo(Y))$ when $(X,Y)\sim\mEqu[^1]$ is the same as sampling $(Al^*(Z),Bo^*(Z))$ when $Z\sim\U[\01^{|u|}]$ -- therefore, $(Al^*(Z),Bo^*(Z))$ is a good pair of messages accepted by the referee with probability at least $1-2\eps-2\sq\eps$.

We will see next that a good pair of messages accepted by the referee defines a good rectangle (Def.~\ref{d_good}); this will imply the lemma, as the rectangle corresponding to the pair of messages $(Al^*(Z),Bo^*(Z))$ under $Z\sim\U[\01^{|u|}]$ is distributed the same way as $\Cl A\times\Cl B$.

Suppose that $(a,b)$ is a good pair of messages accepted by the referee and let $[(a,b)]$ denote the event $[(Al^*(Z),Bo^*(Z))=(a,b)]$.
Then
\m{
\PRr[(X,Y)\unin\01^{n+n}]{(a,b)}{X_u\neq Y_u}
&=\PRr[\mEqu]{(a,b)}{X_u\neq Y_u}\\
&=\PRr[\mEqu]{X_u\neq Y_u}
{(a,b)}\tm\fr{\PR[\mEqu]{(a,b)}}{\PR[\mEqu]{X_u\neq Y_u}}\\
&\le 2\sq\eps\tm\PR[\mEqu]{(a,b)}
,}
as $\PR[\mEqu]{X_u\neq Y_u}=\dr12$.
Similarly,
\m{
\PRr[(X,Y)\unin\01^{n+n}]{(a,b)}{X_u=Y_u} \ge 2(1-\sq\eps)\tm\PR[\mEqu]{(a,b)}
.}
So,
\m{
\PRr[(X,Y)\unin\01^{n+n}]{(a,b)}{X_u\neq Y_u}
\le \fr{\sq\eps}{1-\sq\eps}\tm\PRr[(X,Y)\unin\01^{n+n}]{(a,b)}{X_u=Y_u}
}
and
\m{
\PR[(X,Y)\unin\01^{n+n}]{(a,b)}
&\le \PR{X_u=Y_u}\tm\PRr{(a,b)}{X_u=Y_u}+\PRr{(a,b)}{X_u\neq Y_u}\\
&\le \l(\fr1{2^{|u|}}+\fr{\sq\eps}{1-\sq\eps}\r)\tm\PRr{(a,b)}{X_u=Y_u}
.}
Finally,
\m[m_good-ab]{
\PRr[(X,Y)\unin\01^{n+n}]{X_u=Y_u}{(a,b)}
&=\fr{\PRr{(a,b)}{X_u=Y_u}}{\PR{(a,b)}}\tm\PR{X_u=Y_u}\\
&\ge\fr1{\fr1{2^{|u|}}+\fr{\sq\eps}{1-\sq\eps}}\tm\fr1{2^{|u|}}
\,>\, \fr1{4\sq\eps+2^{-|u|}}\tm\fr1{2^{|u|}}
,}
as $\eps<\dr12$.
The result follows from the definition of good rectangles.
\prfend[\lemref{l_Eq-good}]

We are ready for the main statement of this part.

\crl[crl_Eq]{Let $\Cl P$ be a protocol that solves \Equ\ in \DII[\mEqu,\eps], with $\Delta_\alpha^{u,i}$ and $\Delta_\beta^{u,i}$ as defined earlier.
Then
\m{
\sum_{i=1}^{|u|} \ip{\Delta_\alpha^{u,i}}{\Delta_\beta^{u,i}}
> \fr14\tm\ln\l(\fr1{4\sq\eps+2^{-|u|}}\r) - 2\sq\eps\tm |u|
.}
}

\prfstart
We analyse the quantity
\m{
\E[(\Cl A,\Cl B),\,Z']
{\sum_{i=1}^{|u|}
\delta_{\Cl A}^{u,i}(Z')\tm\delta_{\Cl B}^{u,i}(Z')
}
.}
On the one hand,
\m{
&\hspace{-25pt}
\E[(\Cl A,\Cl B),\,Z']
{\sum_{i=1}^{|u|}
\delta_{\Cl A}^{u,i}(Z')\tm\delta_{\Cl B}^{u,i}(Z')
}\\
&\ge\PR{\txt{$\Cl A\times\Cl B$ is a good rectangle}}
\tm \fr14\tm \ln\l(\fr1{4\sq\eps+2^{-|u|}}\r)\\
&\hspace{35pt}
+\l(1-\PR{\txt{$\Cl A\times\Cl B$ is a good rectangle}}\r)
\tm \Minn{A,B,z}
{\sum_{i=1}^{|u|}
\delta_A^{u,i}(z) \tm\delta_B^{u,i}(z)}\\
&> \l(\fr14-\sq\eps\r)\tm \ln\l(\fr1{4\sq\eps+2^{-|u|}}\r)
-\sq\eps\tm |u|\\
&\ge \fr14\tm\ln\l(\fr1{4\sq\eps+2^{-|u|}}\r) - 2\sq\eps\tm |u|
,}
where the first inequality is \lemref{l_good-cha} and the second one is \lemref{l_Eq-good}.
On the other hand,
\m{
\E[(\Cl A,\Cl B),\,Z']
{\sum_{i=1}^{|u|}
\delta_{\Cl A}^{u,i}(Z')\tm\delta_{\Cl B}^{u,i}(Z')
}
&=\E[Z,\,(\Cl A,\Cl B)]
{\sum_{i=1}^{|u|}
\delta_{\Cl A}^{u,i}(Z)\tm\delta_{\Cl B}^{u,i}(Z)
}\\
&\hspace{-75pt}
=\sum_{i=1}^{|u|} \E[Z\unin\01^{|u|}]
{
\l(
\E[A\sim\alpha^*(Z)]{\delta_{A}^{u,i}(Z)}
\r)\tm\l(
\E[B\sim\beta^*(Z)]{\delta_{B}^{u,i}(Z)}
\r)
}\\
&\hspace{-75pt}
=\sum_{i=1}^{|u|} \E[Z\unin\01^{|u|}]
{\Delta_\alpha^{u,i}(Z)\tm\Delta_\beta^{u,i}(Z)}
\,=\,\sum_{i=1}^{|u|} \ip{\Delta_\alpha^{u,i}}{\Delta_\beta^{u,i}}
,}
where the first equality is \lemref{l_sym}.
\prfend[\crlref{crl_Eq}]

\ssect{Characterising protocols for \EquT}

\lem[l_EqT]{Let $T$ be a \f\delta-biased space for some $\delta>0$ and assume that $\Cl P$ solves $\EquT(X,Y)$ in \DII[\mEquT,\eps].
Then
\m{
&\sum_{i\in u}
\I[X\unin\01^n]{X_i}{X_{u\mset i},Al(X)}
\tm\I[Y\unin\01^n]{Y_i}{Y_{u\mset i},Bo(Y)}\\
&\hspace{25pt}
\in \asOm{\ln\l(\fr1{|T|\tm(\eps+2^{-|u|})}\r)}
-\asO{\l(\sq{|T|\tm\eps} + \delta\r)\tm |u|}
.}
}

\prfstart
From the definition of $\mEquT$ and the correctness assumption it follows that for any $\tau\in T$, if $(X+\tau,Y)\sim\mEqu$, then $\Cl P$ solves $\Equ(X+\tau,Y)$ with error at most
\m{\eps_T\deq|T|\tm(\eps+2^{-|u|}).}

Let $T_u\deq\sett{\tau'}{\tau'_u\in T|_u, \tau'_{[n]\smin u}=\bar0}$ -- in other words, $T_u$ contains the elements of $T$ with bits outside $u$ set to $0$.
To keep the notation simple, assume that $|T_u|=|T|$~\footnotemark, and therefore, $T_u|_u\sbseq\01^{|u|}$ is a \f\delta-biased space.\footnotetext
{
This assumption does not cause any loss of generality:\ without it we would view $T_u$ as a ``multiset''.
}

Observe that for any $\tau\in T$ and the corresponding $\tau'\in T_u$, it holds that $\Equ(X+\tau,Y)\=\Equ(X+\tau',Y)$ and $(X+\tau,Y)\sim\mEqu$ whenever $(X+\tau',Y)\sim\mEqu$.
Accordingly, $\Cl P$ solves $\Equ(X+\tau',Y)$ when $(X+\tau',Y)\sim\mEqu$ with error at most $\eps_T$.

\crlref{crl_Eq} implies that
\m{
\E[\tau'\unin T_u]{
\sum_{i=1}^{|u|} \ip{\Delta_{\alpha,\tau'}^{u,i}}{\Delta_\beta^{u,i}}
}
> \fr14\tm\ln\l(\fr1{4\sq{\eps_T}+2^{-|u|}}\r) - 2\sq{\eps_T}\tm |u|
}
for $\Delta_{\alpha,\tau'}^{u,i}(z)\deq\Delta_\alpha^{u,i}(z\+\tau_u')$ for every $z\in\01^{|u|}$ and $\tau'\in T_u$.
For any $i\in[|u|]$:
\m{
\E[\tau'\unin T_u]
{\ip{\Delta_{\alpha,\tau'}^{u,i}}{\Delta_\beta^{u,i}}}
&= \E[\tau']
{
\sum_{s\sbs[|u|]}
\widehat{\Delta_{\alpha,\tau'}^{u,i}}(s)
\tm\widehat{\Delta_{\beta}^{u,i}}(s)
}\\
&=\sum_{s\sbs[|u|]}
\E[\tau']
{
\widehat{\Delta_{\alpha}^{u,i}}(s)
\tm\chi_s(\tau_u')
\tm\widehat{\Delta_{\beta}^{u,i}}(s)
}\\
&=\sum_{s\sbs[|u|]}
\l(
\widehat{\Delta_{\alpha}^{u,i}}(s)
\tm\widehat{\Delta_{\beta}^{u,i}}(s)
\tm\E[\tau']{\chi_s(\tau_u')}
\r)\\
&\le\widehat{\Delta_{\alpha}^{u,i}}(\emptyset)
\tm\widehat{\Delta_{\beta}^{u,i}}(\emptyset)
+\fr14\tm\Maxx{s\neq\emptyset}{\E[\tau']{\chi_s(\tau_u')}}\\
&\le\widehat{\Delta_{\alpha}^{u,i}}(\emptyset)
\tm\widehat{\Delta_{\beta}^{u,i}}(\emptyset)
+\fr\delta4
,}
where the first two equalities are basic properties of the Fourier transform (see Sect.~\ref{s_prelim}), the first inequality follows from the Parseval's identity and the fact that $\sz{\Delta_{\alpha}^{u,i}(z)},\sz{\Delta_{\beta}^{u,i}(z)}\le\dr12$ for every $z$, and the ultimate step utilises the crucial property of $T_u|_u\sbseq\01^{|u|}$ being a \f\delta-biased space.
So,
\m[m_emp-emp]{
\sum_{i=1}^{|u|}
\widehat{\Delta_{\alpha}^{u,i}}(\emptyset)
\tm\widehat{\Delta_{\beta}^{u,i}}(\emptyset)
> \fr14\tm\ln\l(\fr1{4\sq{\eps_T}+2^{-|u|}}\r)
- \l( 2\sq{\eps_T} + \fr\delta4 \r)\tm |u|
.}

Let us take a closer look at $\widehat{\Delta_{\alpha}^{u,i}}(\emptyset)$.
\m{
\widehat{\Delta_{\alpha}^{u,i}}(\emptyset)
&=\E[Z\unin\01^{|u|}]
{\Delta_{\alpha}^{u,i}(Z)}\\
&=\E[Z]
{
\PRr[X\unin \alpha^*(Z)]
{X_u(i)=Z_{i}}{X_u([i-1])=Z_{[i-1]}}-\fr12
}\\
&=\E[\mac{Z\\A\sim\alpha^*(Z)}]
{
\PRr[X\unin A]
{X_u(i)=Z_{i}}
{X_u([i-1])=Z_{[i-1]}}
- \fr12
}
.}
By the definition of $\alpha^*$ (Def.~\ref{d_pEqu}), the ``chain''
\m{
Z\unin\01^{|u|}
\,\to\,
\Cl A\sim\alpha^*(Z)
\,\to\,
X\unin \Cl A
}
results in the same distribution of $(Z,\Cl A,X)$ as
\m{
X\unin\01^n
\,\to\,
\Cl A=\alpha(X)
\,\to\,
X'\unin \Cl A
\,\to\,
Z=X_u'
.}
Therefore,
\m{
\widehat{\Delta_{\alpha}^{u,i}}(\emptyset)
&= \E[X\unin\01^n]
{
\PRr[X'\unin\alpha(X)]
{X_u(i)=X_u'(i)}
{X_u([i-1])=X_u'([i-1])}
- \fr12
}
.}

Moreover, the marginal distributions of $(\Cl A,X)$ and of $(\Cl A,X')$ are the same:\ we can sample $(X,\Cl A,X')$ by first drawing $\Cl A$ according to its distribution\fn
{
This is the distribution where the probability of $\Cl A=\Cll a$ is proportional to $|\Cll a|$.
}%
, followed by mutually-independent selecting $X\unin \Cl A$ and $X'\unin \Cl A$.
Accordingly,
\m{
\widehat{\Delta_{\alpha}^{u,i}}(\emptyset)
&= \E[\Cl A]
{
\PRr[\mac{X\unin\Cl A\\X'\unin\Cl A}]
{X_u(i)=X_u'(i)}
{X_u([i-1])=X_u'([i-1])}
- \fr12
}\\
&= \E[\mac{\Cl A\\X'\unin\Cl A}]
{
\PRr[X\unin\Cl A]
{X_u(i)=X_u'(i)}
{X_u([i-1])=X_u'([i-1])}
- \fr12
}\\
&\hspace{-25pt}
= \E[\mac{\Cl A'\\X'\unin\Cl A'}]
{
\PRr[\mac{\Cl A\\X\unin\Cl A}]
{X_u(i)=X_u'(i)}
{X_u([i-1])=X_u'([i-1]),\Cl A=\Cl A'}
- \fr12
}
,}
where $\Cl A'$ is distributed identically to $\Cl A$.

Denote $W=(\Cl A,X_u([i-1]))$ and $W'=(\Cl A',X_u'([i-1]))$.
As the marginal distribution of $X$ is uniform, we can apply the second part of \clmref{cl_pred} with respect to $W$ and $X_u(i)$:
\m{
\widehat{\Delta_{\alpha}^{u,i}}(\emptyset)
&= \E[W',X_u'(i)]
{
\PRr[W,X_u(i)]{X_u(i)=X_u'(i)}{W=W'} - \fr12
}\\
&\in\asT{\I{X_u(i)}W}\\
&=\asT{\I{X_u(i)}{\alpha(X),X_u([i-1])}}\\
&=\asT{\I[X\unin\01^n]{X_u(i)}{Al(X),X_u([i-1])}}
.}

Applying similar reasoning to $\widehat{\Delta_{\beta}^{u,i}}(\emptyset)$ and plugging into \bref{m_emp-emp} leads to
\m{
&\sum_{i=1}^{|u|}
\I[X\unin\01^n]{X_u(i)}{Al(X),X_u([i-1])}
\tm\I[Y\unin\01^n]{Y_u(i)}{Bo(Y),Y_u([i-1])}\\
&\hspace{25pt}
\in\asOm{\ln\l(\fr1{\eps_T+2^{-|u|}}\r)}
-\asO{(\sq{\eps_T} + \delta)\tm |u|}
.}
By monotonicity of mutual information,
\m{
&\sum_{i\in u}
\I[X\unin\01^n]{X_i}{Al(X),X_{u\mset i}}
\tm\I[Y\unin\01^n]{Y_i}{Bo(Y),Y_{u\mset i}}\\
&\hspace{25pt}
\in\asOm{\ln\l(\fr1{\eps_T+2^{-|u|}}\r)}
-\asO{(\sq{\eps_T} + \delta)\tm |u|}
,}
as required.
\prfend[\lemref{l_EqT}]

\ssect{Characterising protocols for \tEqT}

\lem[l_tEqT]{For sufficiently large $n$, some $\delta\in\asT{\fr1n}$, any \f\delta-biased space $T$ of size $2^{\aso n}$ and some $\eps\in\asT{\fr1{|T|\tm n^3}}$, any protocol $\Cl P$ that solves $\tEqT(X,Y)$ in \DII[\mtEqT,\eps] satisfies
\m{
\sum_{i=1}^n
\E[u_1]{\I[X\unin\01^n]{X_i}{X_{u_1},Al(X)}}
\tm\E[u_2]{\I[Y\unin\01^n]{Y_i}{Y_{u_2},Bo(Y)}}
>\log n
,}
where $u_1,u_2\unin\chs{[n]\mset i}{\dr{2n}3}$.
}

\prfstart
Suppose that a protocol solves \tEqT\ with respect to \mtEqT\ with error at most $\eps'$, and let $\eps_u'$ be the error that the same protocol makes in solving \EquT\ with respect to \mEquT.

By the definition of the two distributions (Sect.~\ref{sss_Eq}),
\m{
\mtEqT = \E[u\unin\chs{[n]}{\dr n3}]{\mEquT} 
.}
Therefore, $(X,Y)\sim\mtEqT$ can be sampled by first choosing $u\unin\chs{[n]}{\dr n3}$, followed by $(X,Y)\sim\mEquT$.
Then
\m{
\sz{\E[u\unin\chs{[n]}{\dr n3}]{\eps_u'} - \eps'}
& \le \E[u\unin\chs{[n]}{\dr n3}]{\PR[(X,Y)\sim\mEquT]{\tEqT(X,Y) \neq \EquT(X,Y)}}\\
& \le \E[u\unin\chs{[n]}{\dr n3}]{\PR[(X,Y)\sim\wbr\mEquT]{\tEqT(X,Y) \neq \EquT(X,Y)}}
+ 2^{-\asOm n}
,}
where the latter inequality is \clmref{cl_distr_l1}.
As our construction of $T$ is such that $\sz{\tau_1\+\tau_2}\in\fr n2\pm\aso n$ for every $\tau_1\neq\tau_2\in T$ (cf.~\fctref{f_s-bias}), it follows from the Chernoff bound (\fctref{f_Cher}) that
\m{
\PR[u\unin\chs{[n]}{\dr n3}]
{\sz{(\tau_1\+\tau_2)_u}\in\l(\fr{9n}{60},\fr{11n}{60}\r)}
\in 1 - 2^{-\asOm n}
,}
and, on the other hand, it follows by the same \fctref{f_Cher} from the definitions of $\wbr\mEquT$, $\tEqT$ and $\EquT$ that 
\m{
\sz{(\tau_1\+\tau_2)_u}\in\l(\fr{9n}{60},\fr{11n}{60}\r)
~~\Then~~
\PR[(X,Y)\sim\wbr\mEquT]{\tEqT(X,Y) \neq \EquT(X,Y)} \in 2^{-\asOm n}
.}
Accordingly,
\m{
\E[u\unin\chs{[n]}{\dr n3}]{\eps_u'}\le\eps'+2^{-\asOm n}
.}

From \lemref{l_EqT}, our assumption about $|T|$ and the concavity of $\log(\dr1x)$, there exist choices of $\eps$ and $\delta$ in the range given by our statement, so that
\m{
\E[u\unin\chs{[n]}{\dr n3}]
{\sum_{i\in u}
\I[X\unin\01^n]{X_i}{Al(X),X_{u\mset i}}
\tm\I[Y\unin\01^n]{Y_i}{Bo(Y),Y_{u\mset i}}}
\ge2\log n
,}
and therefore for sufficiently large $n$,
\m{
&\hspace{-37pt}
\E[u_1,u_2\unin\chs{[n]}{\dr{2n}3}]
{\sum_{i\in u_1\cap u_2}
\I[X]{X_i}{Al(X),X_{u_1\cap u_2\mset i}}
\tm\I[Y]{Y_i}{Bo(Y),Y_{u_1\cap u_2\mset i}}}\\
&\ge\PR[u_1,u_2\unin\chs{[n]}{\dr{2n}3}]{\sz{u_1\cap u_2}\ge\dr n3}
\tm2>\log n
.}

By the monotonicity of mutual information,
\mal{
\log n
&<\E[u_1,u_2\unin\chs{[n]}{\dr{2n}3}]
{\sum_{i\in u_1\cap u_2}
\I[X]{X_i}{Al(X),X_{u_1\mset i}}
\tm\I[Y]{Y_i}{Bo(Y),Y_{u_2\mset i}}}\\
&\le\sum_{i=1}^n
\E[u_1,u_2\unin\chs{[n]\mset i}{\dr{2n}3}]
{
\I[X]{X_i}{Al(X),X_{u_1}}
\tm\I[Y]{Y_i}{Bo(Y),Y_{u_2}}
}
,}
as required.
\prfend[\lemref{l_tEqT}]

\ssect{Characterising protocols for \ctEqT}

\lem[l_ctEqT]{For sufficiently large $n$, some $\delta\in\asT{\fr1n}$, any \f\delta-biased space $T$ of size $2^{\aso n}$ and some $\eps\in\asT{\fr1{|T|\tm n^3}}$, any protocol $\Cl P$ that solves $\ctEqT(X,Y)$ in \DII[\mctEqT,\eps] satisfies
\m{
\E[i_1,u_1]{\I[X\unin\01^n]{X_{i_1}}{X_{u_1},Al(X)}}
\tm\E[i_2,u_2]{\I[Y\unin\01^n]{Y_{i_2}}{Y_{u_2},Bo(Y)}}
>\fr{\log n}{2n}
,}
where $i_1,i_2\unin{[n]}$, $u_1\unin\chs{[n]\mset {i_1}}{\dr{2n}3}$ and 
$u_2\unin\chs{[n]\mset {i_2}}{\dr{2n}3}$.
}

\prfstart
Suppose that a protocol solves \ctEqT\ with respect to \mctEqT\ with error at most $\eps'$.

By the definition of the input distributions (Sect.~\ref{sss_Eq}),
\m{
\mctEqT = \E[{j\in[n]}]{\mctEqT[^j]} 
.}
Therefore, with probability at least $\dr12$ with respect to $j\unin{[n]}$, the same protocol solves \ctEqT\ with error at most $2\eps'$ with respect to $\mctEqT[^j]$ and -- according to \clmref{cl_distr_l1} -- with error at most $2\eps'+2^{-\asOm n}$ with respect to $\wbr{\mctEqT[^j]}$.
Let $J\sbseq[n]$ be the set of indices $j$ for which the above holds, then $|J|\ge\dr n2$.

Let $j_0\in J$.
It follows by the Chernoff bound (\fctref{f_Cher}) from the definitions of $\wbr{\mctEqT[^j]}$, $\ctEqT$ and $\tEqT$ that 
\m{
\PR[(X,Y)\sim\wbr{\mctEqT[^{j_0}]}]{\ctEqT(X,Y) \neq \tEqT(\sigma_{j_0}(X),Y)} \in 2^{-\asOm n}
,}
and therefore our protocol solves $\tEqT(\sigma_{j_0}(X),Y)$ with error at most $2\eps'+2^{-\asOm n}$ with respect to $(X,Y)\sim\wbr{\mctEqT[^{j_0}]}$, which corresponds to $(\sigma_{j_0}(X),Y)\sim\wbr\mtEqT$.
Via another application of \clmref{cl_distr_l1}, this means that the protocol solves $\tEqT(\sigma_{j_0}(X),Y)$ with error at most $2\eps'+2^{-\asOm n}$ with respect to $(\sigma_{j_0}(X),Y)\sim\mtEqT$.

Accordingly, \lemref{l_tEqT} implies that for some choices of $\eps$ and $\delta$ in the range allowed by our statement the following holds:
\m{
&\E[j\in{[n]}]
{
\sum_{i=1}^n
\E[u_1]{\I[X\unin\01^n]{X_i}{X_{u_1},Al(X)}}
\tm\E[u_2]{\I[Y\unin\01^n]{Y_{\sigma_j(i)}}{Y_{\sigma_j(u_2)},Bo(Y)}}
}\\
&\tb \ge \fr12 \tm \E[j\in J]
{
\sum_{i=1}^n
\E[u_1]{\I[X\unin\01^n]{X_i}{X_{u_1},Al(X)}}
\tm\E[u_2]{\I[Y\unin\01^n]{Y_{\sigma_j(i)}}{Y_{\sigma_j(u_2)},Bo(Y)}}
}\\
&\tb > \fr{\log n}2
,}
where $u_1,u_2\unin\chs{[n]\mset i}{\dr{2n}3}$.
That is,
\m{
\fr{\log n}{2n}
&<\E[i_1,i_2\unin{[n]}]
{
\E[u_1]{\I[X]{X_{i_1}}{X_{u_1},Al(X)}}
\tm\E[u_2]{\I[Y]{Y_{i_2}}{Y_{u_2},Bo(Y)}}
}
,}
where $u_1\unin\chs{[n]\mset {i_1}}{\dr{2n}3}$ and 
$u_2\unin\chs{[n]\mset {i_2}}{\dr{2n}3}$, as required.
\prfend[\lemref{l_ctEqT}]

\crl[crl_ctEqT-low]{There exists a family $\Cl T=T_1,T_2,\ds$, where every $T_i\sbseq\01^i$ can be constructed deterministically in time $\poly(i)$, such that for the corresponding $\ctEqT$ it holds that
\m{
\RIIp(\ctEqT)
\,\ge\, \DII[\mctEqT,\fr13](\ctEqT)
\,\in\, \asOm{\sqrt{\fr n{\log n}}}
.}
}

\prfstart
Let $n$ be sufficiently large, $\delta\in\asT{\fr1n}$ be sufficiently small, $T$ be a \f\delta-biased space of size $\poly(\dr n\delta)$ (as guaranteed by \fctref{f_s-bias}) and $\eps\in\fr1{\poly(n)}$ be sufficiently small, so that \lemref{l_ctEqT} guarantees that for any protocol $\Cl P$ solving \ctEqT\ in \DII[\mctEqT,\eps] it holds that
\m{
\E[i_1,u_1]{\I[X\unin\01^n]{X_{i_1}}{X_{u_1},Al(X)}}
\tm\E[i_2,u_2]{\I[Y\unin\01^n]{Y_{i_2}}{Y_{u_2},Bo(Y)}}
>\fr{\log n}{2n}
.}

Without loss of generality, assume that
\m{
\E[i_1,u_1]{\I[X\unin\01^n]{X_{i_1}}{X_{u_1},Al(X)}}
>\sq{\fr{\log n}{2n}}
}
for $i_1\unin{[n]}$ and $u_1\unin\chs{[n]\mset {i_1}}{\dr{2n}3}$, then
\m{
\exists\, u\in\chs{[n]}{\dr{2n}3}:
\sum_{i\nin u}{\I[X\unin\01^n]{X_i}{X_{u},Al(X)}}
>\fr n3\tm\sq{\fr{\log n}{2n}}
,}
and therefore the complexity of $\Cl P$ is at least
\m{
\Ii[X\unin\01^n]{Al(X)}X{X_{u}} > \fr{\sq{n\tm\log n}}6
.}

If, on the other hand, a protocol solves \ctEqT\ in \DII[\mctEqT,\fr13], then repeated $k$ times in parallel for a sufficient $k\in\asO{\log n}$, it would solve \ctEqT\ with error at most $\eps$.
\prfend[\crlref{crl_ctEqT-low}]

\sect[s_land]{Conclusion}

From \crlref[crl_ctEqT-low]{crl_ctEqT-upper}:
\crl[crl_ctEqT-both]{There exists a family $\Cl T=T_1,T_2,\ds$, where every $T_i\sbseq\01^i$ can be constructed deterministically in time $\poly(i)$ and for the corresponding $\ctEqT$ it holds that
\m{
\QII(\ctEqT)
\,\in\, \asO{(\log n)^2}
\tb\txt{and}\tb
\RIIp(\ctEqT)
\,\in\, \asOm{\sq{\fr n{\log n}}}
.}
}

\para{The landscape of quantum superiority and further questions.}
One of the main questions related to quantum communication complexity is ``\e{When can quantum outperform classical?}'' -- formally, for which pairs of quantum and classical communication models the former is super-polynomially\fn
{
All known super-polynomial separations are, in fact, exponential.
}
more efficient than the latter in solving a specific problem.

There are three main \e{types} of communication problems used for model separations:\ \e{functions}, \e{total functions} and \e{relations}.
Functions -- probably, the most natural class of communication problems -- are a special case of relations.
Total functions are a restricted special case of functions, where the support is required to be the product set of the players' individual sets of input.\fn
{
To emphasise the distinction from total functions in the context of communication complexity, the term \e{partial functions} is often used to address the unrestricted functions.
}
There are known cases where a quantum communication complexity class can be separated from a classical one via a relation, while a functional separation is provably impossible (see~\cite{A04_Lim,GRW08_Sim}).

The history of (super-polynomial) separations that showed advantage of quantum communication can be briefly outlined as follows.
\itstart
\item In 1999 Raz~\cite{R99_Ex} demonstrated a \e{function} that had an efficient \e{quantum two-way protocol}, but no efficient \e{classical two-way protocol}.
\item In 2001 Buhrman, Cleve, Watrous and de Wolf~\cite{BCWW01_Qua} demonstrated a \e{total function} (namely, the equality) that had an efficient \e{quantum simultaneous-messages protocol without shared randomness}, but no efficient \e{classical simultaneous-messages protocol without shared randomness}.
\item In 2004 Bar-Yossef, Jayram and Kerenidis~\cite{BJK04_Exp} demonstrated a \e{relation} that had an efficient \e{quantum simultaneous-messages protocol without shared randomness}, but no efficient \e{classical one-way protocol}.
\item In 2007 in a joint work with Kempe, Kerenidis, Raz and de Wolf~\cite{GKKRW08_Ex} a \e{function} was demonstrated, that had an efficient \e{quantum one-way protocol}, but no efficient \e{classical one-way protocol}.
\item In 2008 a \e{relation} was demonstrated~\cite{G08_Cla} with an efficient \e{quantum one-way protocol}, but no efficient \e{classical two-way protocol}.
\item In 2010 Klartag and Regev~\cite{KR11_Qua} demonstrated a \e{function} with an efficient \e{quantum one-way protocol}, but no efficient \e{classical two-way protocol}.
\item In 2016 a \e{function} was demonstrated~\cite{G16_En} with an efficient \e{quantum simultaneous-messages protocol with entanglement}, but no efficient \e{classical two-way protocol}.
\item This work presents a \e{function} with an efficient \e{quantum simultaneous-messages protocol without shared randomness}, but no efficient \e{classical simultaneous-messages protocol with shared randomness}.
\itend

Is it the case that ``everything separable'' has already been discovered -- in other words, that for the pairs of a quantum and a classical model, where we do not yet have an example of quantum superiority, such examples do not exist?
Our current knowledge of ``limitations to separability'' is very limited:\ in particular, virtually nothing is known in this respect regarding the models considered in this work.

To summarise what is known and what is still missing, let us consider the three ``canonical'' randomised models: two-way (\R), one-way (\RI) and SMP (\RIIp), and add to our picture the ``purposely weakened'' SMP (\RII).
We are interested in their ``strength relationship'' with the quantum counterparts -- both the \e{closest} (e.g., \R\ vs.\ \Q) and ``topologically'' \e{weaker} (e.g., \R\ vs.\ \QII).
\itstart
\item If we only allow \e{functions} and only consider the \e{closest pairs}, then our knowledge has been completed by this work:
\cent{
\begin{tabular}{ c c c c c c c }
\RII    & $<$ & \RIIp    & $<$ & \RI      & $<$ & \R       \\
$\wedge$ &     & $\wedge$ &     & $\wedge$ &     & $\wedge$ \\
\QII    & $<$ & \QIIp    & $<$ & \QI      & $<$ & \Q
\end{tabular}
}
(we have just seen that $\QIIp>\RIIp$, the rest has been known for some time).

\item As for \e{``diagonal'' relationship} via \e{functions}, it has been known that \QI\ can be stronger than \R\ and we have just seen that \QII\ can be stronger than \RIIp.
\quest[q_partcl]{Can some of $\set{\QII,\QIIp}$ be stronger than some of $\set{\RI,\R}$ with respect to a function?~\fn
{
Although not directly related to quantum superiority, nonetheless an interesting question is:
Can \RIIp\ be stronger than \QII\ with respect to a function?
For relations this possibility has been demonstrated in~\cite{GKRW09_Bo}.
}
}

\item If we allow \e{relational problems}, then one additional \e{``diagonal''} separation is known:\ \QII\ can be stronger than \RI.
\quest{Can \QII\ or \QIIp\ be stronger than \R\ with respect to a relation?}
As we mentioned earlier, looking for \e{the weakest quantum model that can outperform \R} and for \e{the strongest classical model that can be outperformed by \QII} are, probably, the two most natural approaches towards understanding the strength and the limits of quantum communication.
Ultimately, we would like the two approaches to ``meet'' -- that is, to find a communication problem (even a relational one), \e{easy for \QII\ but hard for \R}.

\item For the case of \e{total functions} our current lack of understanding is almost perfect.
We know nothing about \e{``diagonal'' relationship} and nearly nothing about the \e{closest pairs}:
\cent{
\begin{tabular}{ c c c c c c c }
\RII    & $<$ & \RIIp    & $<$ & \RI      & $<$ & \R  \\
$\wedge$ &     & $\red?$  &     & $\red?$  &     & $\red?$ \\
\QII    & $<$ & \QIIp    & $<$ & \QI      & $<$ & \Q
\end{tabular}
}
\quest{In the case of total functions, can \QIIp\ be stronger than \RIIp?
How about \QI\ vs.\ \RI?
\Q\ vs.\ \R?\\
Can \QII, \QIIp\ or \QI\ be stronger than \R?
Can \QII\ be stronger than \RIIp\ or \RI?
}
\itend

Lastly, we would like to mention
\quest{What is the complexity of our $\ctEqT$ in the model of \e{classical SMP with shared entanglement (\RIIe)}?}
If it has an efficient solution, that would imply a functional separation between \RIIe\ and \RIIp, which we do not have yet (a relational separation is known); if, on the other hand, $\ctEqT$ is hard for \RIIe, that would imply the possibility of qualitative advantage of \QII\ over \RIIe, which is currently not known even for relational problems.

\sect*{Acknowledgements}

I am very grateful to Pavel Pudl\'ak, Ronald de Wolf and Thomas Vidick for helpful discussions at various stages of this work, and to Alexander Razborov for finding a mistake in the initial version of the proof of \clmref{cl_optchain}.

A number of very useful comments have been received from anonymous reviewers.
In particular, one of the suggestions has led to improving the lower bound on $\RIIp(\ctEqT)$ from ``$\asOm{\dr{\sq n}{\log n}}$'' to the current ``$\asOm{\sq{\dr n{\log n}}}$''.

\newcommand{\etalchar}[1]{$^{#1}$}

\end{document}